\documentclass[twocolumn]{aastex631}

\usepackage{mathtools, multirow, tabularx, threeparttable}
\usepackage{CJK}
\usepackage[utf8]{inputenc}
\usepackage{etoolbox}

\usepackage[multiple]{footmisc}
\newcommand{\J}[1]{\textbf{\textcolor{teal}{#1}}} 

\received{}
\revised{}
\accepted{}

\shorttitle{}
\shortauthors{}

\graphicspath{{./}{figures/}}

\begin{document}

\title{Mid-Infrared Dust Evolution and Late-time Circumstellar Medium Interaction in SN~2017eaw}

\correspondingauthor{Jeniveve Pearson}
\email{jenivevepearson@arizona.edu}

\newcommand{\LCO}{\affiliation{Las Cumbres Observatory, 6740 Cortona Drive, Suite 102, Goleta, CA 93117-5575, USA}}
\newcommand{\UCSB}{\affiliation{Department of Physics, University of California, Santa Barbara, CA 93106-9530, USA}}
\newcommand{\KITP}{\affiliation{Kavli Institute for Theoretical Physics, University of California, Santa Barbara, CA 93106-4030, USA}}
\newcommand{\UCD}{\affiliation{Department of Physics and Astronomy, University of California, Davis, 1 Shields Avenue, Davis, CA 95616-5270, USA}}
\newcommand{\WIS}{\affiliation{Department of Particle Physics and Astrophysics, Weizmann Institute of Science, 76100 Rehovot, Israel}}
\newcommand{\OKC}{\affiliation{Oskar Klein Centre, Department of Astronomy, Stockholm University, Albanova University Centre, SE-106 91 Stockholm, Sweden}}
\newcommand{\OAPD}{\affiliation{INAF-Osservatorio Astronomico di Padova, Vicolo dell'Osservatorio 5, I-35122 Padova, Italy}}
\newcommand{\UniPd}{\affiliation{Dipartimento di Fisica e Astronomia ``G. Galilei'', Universit\`a degli studi di Padova Vicolo dell'Osservatorio 3, I-35122 Padova, Italy}}
\newcommand{\CaltechCahill}{\affiliation{Cahill Center for Astronomy and Astrophysics, California Institute of Technology, Mail Code 249-17, Pasadena, CA 91125, USA}}
\newcommand{\Caltech}{\affiliation{Department of Astronomy and Astrophysics, California Institute of Technology, Pasadena, CA 91125, USA}}
\newcommand{\GSFC}{\affiliation{Astrophysics Science Division, NASA Goddard Space Flight Center, Mail Code 661, Greenbelt, MD 20771, USA}}
\newcommand{\UMD}{\affiliation{Joint Space-Science Institute, University of Maryland, College Park, MD 20742, USA}}
\newcommand{\UCB}{\affiliation{Department of Astronomy, University of California, Berkeley, CA 94720-3411, USA}}
\newcommand{\TTU}{\affiliation{Department of Physics, Texas Tech University, Box 41051, Lubbock, TX 79409-1051, USA}}
\newcommand{\STScI}{\affiliation{Space Telescope Science Institute, 3700 San Martin Drive, Baltimore, MD 21218-2410, USA}}
\newcommand{\UT}{\affiliation{University of Texas at Austin, 1 University Station C1400, Austin, TX 78712-0259, USA}}
\newcommand{\IoA}{\affiliation{Institute of Astronomy, University of Cambridge, Madingley Road, Cambridge CB3 0HA, UK}}
\newcommand{\QUB}{\affiliation{Astrophysics Research Centre, School of Mathematics and Physics, Queen's University Belfast, Belfast BT7 1NN, UK}}
\newcommand{\IPAC}{\affiliation{IPAC, Mail Code 100-22, Caltech, 1200 E. California Blvd., Pasadena, CA 91125, USA}}
\newcommand{\JPL}{\affiliation{Jet Propulsion Laboratory, California Institute of Technology, 4800 Oak Grove Dr, Pasadena, CA 91109, USA}}
\newcommand{\Southampton}{\affiliation{Department of Physics and Astronomy, University of Southampton, Southampton SO17 1BJ, UK}}
\newcommand{\LANL}{\affiliation{Space and Remote Sensing, MS B244, Los Alamos National Laboratory, Los Alamos, NM 87545, USA}}
\newcommand{\Tsinghua}{\affiliation{Physics Department and Tsinghua Center for Astrophysics, Tsinghua University, Beijing, 100084, People's Republic of China}}
\newcommand{\NAOC}{\affiliation{National Astronomical Observatory of China, Chinese Academy of Sciences, Beijing, 100012, People's Republic of China}}
\newcommand{\Itagaki}{\affiliation{Itagaki Astronomical Observatory, Yamagata 990-2492, Japan}}
\newcommand{\Einstein}{\altaffiliation{Einstein Fellow}}
\newcommand{\Hubble}{\altaffiliation{Hubble Fellow}}
\newcommand{\CfA}{\affiliation{Center for Astrophysics \textbar{} Harvard \& Smithsonian, 60 Garden Street, Cambridge, MA 02138-1516, USA}}
\newcommand{\UA}{\affiliation{Steward Observatory, University of Arizona, 933 North Cherry Avenue, Tucson, AZ 85721-0065, USA}}
\newcommand{\MPIA}{\affiliation{Max-Planck-Institut f\"ur Astrophysik, Karl-Schwarzschild-Stra\ss{}e 1, D-85748 Garching, Germany}}
\newcommand{\DSFP}{\altaffiliation{LSSTC Data Science Fellow}}
\newcommand{\HCO}{\affiliation{Harvard College Observatory, 60 Garden Street, Cambridge, MA 02138-1516, USA}}
\newcommand{\Carnegie}{\affiliation{Observatories of the Carnegie Institute for Science, 813 Santa Barbara Street, Pasadena, CA 91101-1232, USA}}
\newcommand{\TAU}{\affiliation{School of Physics and Astronomy, Tel Aviv University, Tel Aviv 69978, Israel}}
\newcommand{\Edinburgh}{\affiliation{Institute for Astronomy, University of Edinburgh, Royal Observatory, Blackford Hill EH9 3HJ, UK}}
\newcommand{\Birmingham}{\affiliation{Birmingham Institute for Gravitational Wave Astronomy and School of Physics and Astronomy, University of Birmingham, Birmingham B15 2TT, UK}}
\newcommand{\Bath}{\affiliation{Department of Physics, University of Bath, Claverton Down, Bath BA2 7AY, UK}}
\newcommand{\CTIO}{\affiliation{Cerro Tololo Inter-American Observatory, National Optical Astronomy Observatory, Casilla 603, La Serena, Chile}}
\newcommand{\Potsdam}{\affiliation{Institut f\"ur Physik und Astronomie, Universit\"at Potsdam, Haus 28, Karl-Liebknecht-Str. 24/25, D-14476 Potsdam-Golm, Germany}}
\newcommand{\INPE}{\affiliation{Instituto Nacional de Pesquisas Espaciais, Avenida dos Astronautas 1758, 12227-010, S\~ao Jos\'e dos Campos -- SP, Brazil}}
\newcommand{\UNC}{\affiliation{Department of Physics and Astronomy, University of North Carolina, 120 East Cameron Avenue, Chapel Hill, NC 27599, USA}}
\newcommand{\Ohio}{\affiliation{Astrophysical Institute, Department of Physics and Astronomy, 251B Clippinger Lab, Ohio University, Athens, OH 45701-2942, USA}}
\newcommand{\AAS}{\affiliation{American Astronomical Society, 1667 K~Street NW, Suite 800, Washington, DC 20006-1681, USA}}
\newcommand{\MMT}{\affiliation{MMT and Steward Observatories, University of Arizona, 933 North Cherry Avenue, Tucson, AZ 85721-0065, USA}}
\newcommand{\Geneva}{\affiliation{ISDC, Department of Astronomy, University of Geneva, Chemin d'\'Ecogia, 16 CH-1290 Versoix, Switzerland}}
\newcommand{\IUCAA}{\affiliation{Inter-University Center for Astronomy and Astrophysics, Post Bag 4, Ganeshkhind, Pune, Maharashtra 411007, India}}
\newcommand{\CMU}{\affiliation{Department of Physics, Carnegie Mellon University, 5000 Forbes Avenue, Pittsburgh, PA 15213-3815, USA}}
\newcommand{\NAOJ}{\affiliation{Division of Science, National Astronomical Observatory of Japan, 2-21-1 Osawa, Mitaka, Tokyo 181-8588, Japan}}
\newcommand{\IfA}{\affiliation{Institute for Astronomy, University of Hawai`i, 2680 Woodlawn Drive, Honolulu, HI 96822-1839, USA}}
\newcommand{\UCSC}{\affiliation{Department of Astronomy and Astrophysics, University of California, Santa Cruz, CA 95064-1077, USA}}
\newcommand{\Purdue}{\affiliation{Department of Physics and Astronomy, Purdue University, 525 Northwestern Avenue, West Lafayette, IN 47907-2036, USA}}
\newcommand{\Princeton}{\affiliation{Department of Astrophysical Sciences, Princeton University, 4 Ivy Lane, Princeton, NJ 08540-7219, USA}}
\newcommand{\Moore}{\affiliation{Gordon and Betty Moore Foundation, 1661 Page Mill Road, Palo Alto, CA 94304-1209, USA}}
\newcommand{\Durham}{\affiliation{Department of Physics, Durham University, South Road, Durham, DH1 3LE, UK}}
\newcommand{\JHU}{\affiliation{Department of Physics and Astronomy, The Johns Hopkins University, 3400 North Charles Street, Baltimore, MD 21218, USA}}
\newcommand{\Toronto}{\affiliation{David A.\ Dunlap Department of Astronomy and Astrophysics, University of Toronto,\\ 50 St.\ George Street, Toronto, Ontario, M5S 3H4 Canada}}
\newcommand{\Duke}{\affiliation{Department of Physics, Duke University, Campus Box 90305, Durham, NC 27708, USA}}
\newcommand{\NCU}{\affiliation{Graduate Institute of Astronomy, National Central University, 300 Jhongda Road, 32001 Jhongli, Taiwan}}
\newcommand{\Columbia}{\affiliation{Department of Physics and Columbia Astrophysics Laboratory, Columbia University, Pupin Hall, New York, NY 10027, USA}}
\newcommand{\Flatiron}{\affiliation{Center for Computational Astrophysics, Flatiron Institute, 162 5th Avenue, New York, NY 10010-5902, USA}}
\newcommand{\CIERA}{\affiliation{Center for Interdisciplinary Exploration and Research in Astrophysics (CIERA), 1800 Sherman Ave., Evanston, IL 60201, USA}}
\newcommand{\NU}{\affiliation{Department of Physics and Astronomy, Northwestern University, 2145 Sheridan Road, Evanston, IL 60208, USA}}
\newcommand{\SkAI}{\affiliation{NSF-Simons AI Institute for the Sky (SkAI), 172 E. Chestnut St., Chicago, IL 60611, USA}}
\newcommand{\GeminiNorth}{\affiliation{Gemini Observatory, 670 North A`ohoku Place, Hilo, HI 96720-2700, USA}}
\newcommand{\Keck}{\affiliation{W.~M.~Keck Observatory, 65-1120 M\=amalahoa Highway, Kamuela, HI 96743-8431, USA}}
\newcommand{\UW}{\affiliation{Department of Astronomy, University of Washington, 3910 15th Avenue NE, Seattle, WA 98195-0002, USA}}
\newcommand{\DiRAC}{\altaffiliation{DiRAC Fellow}}
\newcommand{\USask}{\affiliation{Department of Physics \& Engineering Physics, University of Saskatchewan, 116 Science Place, Saskatoon, SK S7N 5E2, Canada}}
\newcommand{\Thacher}{\affiliation{Thacher School, 5025 Thacher Road, Ojai, CA 93023-8304, USA}}
\newcommand{\Rutgers}{\affiliation{Department of Physics and Astronomy, Rutgers, the State University of New Jersey,\\136 Frelinghuysen Road, Piscataway, NJ 08854-8019, USA}}
\newcommand{\FSU}{\affiliation{Department of Physics, Florida State University, 77 Chieftan Way, Tallahassee, FL 32306-4350, USA}}
\newcommand{\Melbourne}{\affiliation{School of Physics, The University of Melbourne, Parkville, VIC 3010, Australia}}
\newcommand{\ASTROthreeD}{\affiliation{ARC Centre of Excellence for All Sky Astrophysics in 3 Dimensions (ASTRO 3D)}}
\newcommand{\Stromlo}{\affiliation{Mt.\ Stromlo Observatory, The Research School of Astronomy and Astrophysics, Australian National University, ACT 2601, Australia}}
\newcommand{\NCPAS}{\affiliation{National Centre for the Public Awareness of Science, Australian National University, Canberra, ACT 2611, Australia}}
\newcommand{\TAMU}{\affiliation{Department of Physics and Astronomy, Texas A\&M University, 4242 TAMU, College Station, TX 77843, USA}}
\newcommand{\Mitchell}{\affiliation{George P.\ and Cynthia Woods Mitchell Institute for Fundamental Physics \& Astronomy, College Station, TX 77843, USA}}
\newcommand{\ESO}{\affiliation{European Southern Observatory, Alonso de C\'ordova 3107, Casilla 19, Santiago, Chile}}
\newcommand{\MAS}{\affiliation{Millennium Institute of Astrophysics MAS, Nuncio Monsenor Sotero Sanz 100, Off.
104, Providencia, Santiago, Chile}}
\newcommand{\ICE}{\affiliation{Institute of Space Sciences (ICE, CSIC), Campus UAB, Carrer de Can Magrans, s/n, E-08193 Barcelona, Spain}}
\newcommand{\IEEC}{\affiliation{Institut d'Estudis Espacials de Catalunya, Gran Capit\`a, 2-4, Edifici Nexus, Desp.\ 201, E-08034 Barcelona, Spain}}
\newcommand{\Warwick}{\affiliation{Department of Physics, University of Warwick, Gibbet Hill Road, Coventry CV4 7AL, UK}}
\newcommand{\Macquarie}{\affiliation{School of Mathematical and Physical Sciences, Macquarie University, NSW 2109, Australia}}
\newcommand{\AAARC}{\affiliation{Astronomy, Astrophysics and Astrophotonics Research Centre, Macquarie University, Sydney, NSW 2109, Australia}}
\newcommand{\Capodimonte}{\affiliation{INAF - Capodimonte Astronomical Observatory, Salita Moiariello 16, I-80131 Napoli, Italy}}
\newcommand{\INFNNapoli}{\affiliation{INFN - Napoli, Strada Comunale Cinthia, I-80126 Napoli, Italy}}
\newcommand{\ICRANet}{\affiliation{ICRANet, Piazza della Repubblica 10, I-65122 Pescara, Italy}}
\newcommand{\MSU}{\affiliation{Center for Data Intensive and Time Domain Astronomy, Department of Physics and Astronomy,\\Michigan State University, East Lansing, MI 48824, USA}}
\newcommand{\SETI}{\affiliation{SETI Institute,
339 Bernardo Ave, Suite 200, Mountain View, CA 94043, USA}}
\newcommand{\IAIFI}{\affiliation{The NSF AI Institute for Artificial Intelligence and Fundamental Interactions}}
\newcommand{\ANUC}{\affiliation{Department of Astronomy, AlbaNova University Center, Stockholm University, SE-10691 Stockholm, Sweden}}
\newcommand{\UVA}{\affiliation{Department of Astronomy, University of Virginia, Charlottesville, VA 22904, USA}}
\newcommand{\THCA}{\affiliation{Physics Department and Tsinghua Center for Astrophysics (THCA), Tsinghua University, Beijing, 100084, People's Republic of China}}
\newcommand{\NARIT}{\affiliation{National Astronomical Research Institute of Thailand (NARIT), Don Kaeo, Mae Rim District, Chiang Mai 50180, Thailand}}
\newcommand{\HamObs}{\affiliation{Hamburger Sternwarte, Gojenbergsweg 112, 21029 Hamburg, Germany}}
\newcommand{\VaTech}{\affiliation{Department of Physics, Virginia Tech, 850 West Campus Drive, Blacksburg VA, 24061, USA}}
\newcommand{\Yunnan}{\affiliation{Yunnan Observatories, Chinese Academy of Sciences, Kunming 650216, P.R. China}}
\newcommand{\KLSECO}{\affiliation{Key Laboratory for the Structure and Evolution of Celestial Objects, Chinese Academy of Sciences, Kunming 650216, P.R. China}}
\newcommand{\ICS}{\affiliation{International Centre of Supernovae, Yunnan Key Laboratory, Kunming 650216, P.R. China}}
\newcommand{\FINCA}{\affiliation{Finnish Centre for Astronomy with ESO (FINCA), FI-20014 University of Turku, Finland}}
\newcommand{\Tuorla}{\affiliation{Tuorla Observatory, Department of Physics and Astronomy, FI-20014 University of Turku, Finland}}
\newcommand{\LAM}{\affiliation{Aix-Marseille Univ, CNRS, CNES, LAM, 13388 Marseille, France}}
\newcommand{\IAC}{\affiliation{Instituto de Astrof\'isica de Canarias, E-38205 La Laguna, Tenerife, Spain}}
\newcommand{\LPNHE}{\affiliation{LPNHE, (CNRS/IN2P3, Sorbonne Universit\'e, Universit\'e Paris Cit\'e), Laboratoire de Physique Nucl\'eaire et de Hautes \'Energies, 75005, Paris, France}}
\newcommand{\UniLag}{\affiliation{Universidad de La Laguna, Dept. Astrof\'isica, E-38206 La Laguna, Tenerife, Spain}}
\newcommand{\OU}{\affiliation{Homer L. Dodge Department of Physics and Astronomy, University of Oklahoma, 440 W. Brooks, Norman, OK 73019-2061, USA}}
\newcommand{\PSI}{\affiliation{Planetary Science Institute, 1700 East Fort Lowell Road, Suite 106, Tucson, AZ 85719-2395, USA}}
\newcommand{\TUM}{\affiliation{Technische Universit\"at M\"unchen, TUM School of Natural Sciences, Physik-Department, James-Franck-Stra\ss{}e 1, 85748 Garching, Germany}}
\newcommand{\UWarsaw}{\affiliation{Astronomical Observatory, University of Warsaw, Al. Ujazdowskie 4, 00-478 Warszawa, Poland}}
\newcommand{\Trinity}{\affiliation{School of Physics, Trinity College Dublin, The University of Dublin, Dublin
2, Ireland}}
\newcommand{\UNAB}{\affiliation{Departamento de Ciencias F\'isicas, Facultad de Ciencias Exactas, Universidad Andr\'es Bello, Fern\'andez Concha 700, Las Condes,
Santiago, Chile}}
\newcommand{\CCAPP}{\affiliation{Center for Cosmology and Astroparticle Physics, The Ohio State University, 191 West Woodruff Ave, Columbus, OH, USA}}
\newcommand{\OSU}{\affiliation{Department of Astronomy, The Ohio State University, 140 West 18th Avenue, Columbus, OH, USA}}
\newcommand{\LCOactual}{\affiliation{Las Campanas Observatory, Carnegie Observatories, Casilla 601, La
Serena, Chile}}
\newcommand{\UCSD}{\affiliation{Department of Astronomy \& Astrophysics, University of California, San Diego, 9500 Gilman Drive, MC 0424, La Jolla, CA 92093-0424, USA}}
\newcommand{\Liverpool}{\affiliation{Astrophysics Research Institute, Liverpool John Moores University, 146 Brownlow Hill, Liverpool, L3 5RF, UK}} \author[0000-0002-0744-0047]{Jeniveve Pearson}
\UA
\author[0000-0001-8073-8731]{Bhagya Subrayan}
\UA
\author[0000-0003-4102-380X]{David J. Sand}
\UA
\author[0000-0003-0123-0062]{Jennifer E. Andrews}
\GeminiNorth
\author[0000-0003-4666-4606]{Emma R. Beasor}
\Liverpool
\author[0000-0002-4924-444X]{K. Azalee Bostroem}
\altaffiliation{LSST-DA Catalyst Fellow}
\UA
\author[0000-0002-7937-6371]{Yize Dong \begin{CJK*}{UTF8}{gbsn}(董一泽)\end{CJK*}}
\CfA
\author[0000-0003-2744-4755]{Emily Hoang}
\UCD
\author[0000-0002-0832-2974]{Griffin Hosseinzadeh}
\UCSD
\author[0000-0002-9454-1742]{Brian Hsu}
\UA
\author[0000-0002-3934-2644]{Wynn Jacobson-Gal{\'a}n}
\altaffiliation{NASA Hubble Fellow}
\Caltech
\author[0000-0003-0549-3281]{Daryl Janzen}
\USask
\author[0000-0001-5754-4007]{Jacob Jencson}
\IPAC
\author[0000-0001-8738-6011]{Saurabh W.\ Jha}
\Rutgers
\author[0000-0002-5740-7747]{Charles D. Kilpatrick}
\NU \CIERA
\author[0000-0003-3108-1328]{Lindsey A. Kwok}
\CIERA
\author[0000-0002-7866-4531]{Chang Liu}
\NU \CIERA
\author[0000-0001-9589-3793]{M.~J. Lundquist}
\Keck
\author[0009-0008-9693-4348]{Darshana Mehta}
\UCD
\author[0000-0001-9515-478X]{Adam A. Miller}
\NU \CIERA \SkAI
\author[0000-0002-7352-7845]{Aravind P.\ Ravi}
\UCD
\author[0000-0002-5683-2389]{Nabeel Rehemtulla}
\NU \CIERA \SkAI
\author[0000-0002-7015-3446]{Nicol\'as Meza Retamal}
\UCD
\author[0000-0002-4022-1874]{Manisha Shrestha}
\UA
\author[0000-0001-5510-2424]{Nathan Smith}
\UA
\author[0000-0001-8818-0795]{Stefano Valenti}
\UCD
\author[0000-0003-1432-7744]{Lily Whitler}
\UA 
\begin{abstract}
We present JWST/MIRI and complementary ground-based near-infrared observations of the Type II SN~2017eaw taken 6 years post-explosion. SN~2017eaw is still detected out to 25 $\mu$m and there is minimal evolution in the mid-infrared spectral energy distribution (SED) between the newly acquired JWST/MIRI observations and those taken a year earlier. 
Modeling of the mid-infrared SED reveals a cool $\sim$160~K dust component of $5.5\times10^{-4}\ \mathrm{M}_\odot$ and a hot $\sim$1700~K component of $5.4\times10^{-8}\ \mathrm{M}_\odot$ both composed of silicate dust. 
Notably there is no evidence of temperature or mass evolution in the cool dust component in the year between JWST observations.
We also present new and archival HST and ground-based ultraviolet (UV) and optical observations which reveal reduced but continued circumstellar medium (CSM)-ejecta interaction at $>$2000 days post-explosion. 
The UV and mid-infrared emission show similar decline rates, suggesting both probe the interface between the ejecta and CSM. Given this, the continued existence of boxy H$\alpha$ emission in the nebular spectra, the low inferred optical depth of the dust, and the lack of temperature and mass evolution, we suggest that the cool dust component in SN~2017eaw may be primarily due to pre-existing dust rather than newly-formed dust in the ejecta or cold dense shell.

\end{abstract}


\section{Introduction} \label{sec:intro}

Observations of high redshift galaxies have revealed significant amounts of dust in the early universe \citep{Marrone18, Hashimoto19, Witstok23, Markov24, Nanni2025}. 
The majority of this dust is likely associated with core-collapse supernovae \citep[CCSNe;][]{Morgan03_earlygal, Maiolino04, Gall11, Schneider24}. 
Models of high redshift supernovae (SNe) and star formation rates indicate that SNe would need to produce between $0.1-1$ M$_\odot$ of dust per SN \citep{Todini01, Sarangi18, Schneider24} and observations of nearby SN remnants have revealed dust masses within this range \citep{Dunne03, Morgan03_keplers, Rho08, DeLooze17, Chastenet22, Priestley22}. However, the vast majority of near- and mid-infrared studies of nearby SNe undertaken prior to JWST have revealed significantly lower dust masses \citep[$\lesssim$10$^{-2}$ $M_{\odot}$;][]{Gall11, Szalai13,Szalai19_spitzer, Tinyanont16}. These previous studies therefore suggest that infrared observations of CCSNe in the decades after explosion may be missing a significant portion of the dust.  

Dust formation in SNe likely occurs in the expanding ejecta interior to the reverse shock and/or in a cold dense shell between the forward and reverse shock created by the interaction between the forward shock and surrounding dense circumstellar material (CSM) \citep[e.g.][]{Pozzo04, Mattila08, smith08jc, Smith09}. In the decades to centuries following the SN, some of the dust in the interior ejecta will be destroyed by interaction with the reverse shock; however, some percentage is expected to survive this interaction and facilitate further dust formation {\citep{Bianchi2007, Nozawa2007, Silvia2010}}. 
There are some indications that the majority of newly formed dust in the first years post-explosion in non-interacting {hydrogen-rich CCSNe (e.g. type IIP/L SNe)} is interior and optically thick and thus only visible in the spectral line profiles and not in infrared images. Studies modeling the optical nebular spectra of {non-interacting H-rich SNe} recover higher dust masses than indicated by infrared photometry alone \citep{Niculescu-Duvaz2022, Zsiros24}. Further, the significant dust mass in SN~1987A is only observable in the far-IR and sub-millimeter \citep{Bouchet06, Matsuura11, Indebetouw14, Matsuura15, Cigan19}. Thus even in the case where the dust is optically thin, any newly-formed dust may be too cold to detect in the near/mid-infrared. 

Measuring dust formation in SNe is further complicated by the presence of pre-existing dust in the CSM. Dust directly around the progenitor is destroyed immediately following the explosion but pre-SN dust can survive at further distances. This pre-existing dust can be formed within the CSM in the stellar winds and/or binary interactions of massive stars. This dust is warmed by the SN explosion and subsequent ejecta-CSM interaction, becoming visible in the infrared \citep[e.g.][]{Sugerman03, Kotak09, Fox10}. Dust in nearby SNe is likely both pre-existing in the CSM and created in the ejecta.  
However, only a handful of SNe have the multi-epoch, multi-wavelength measurements crucial to disentangle the origin of the dust and therefore constrain the timeline of CCSNe dust formation. 

The launch of JWST has ushered in  a new era of SN dust studies. JWST's sensitivity, resolution and wavelength coverage has allowed for observations that probe both hot and cool dust around the SNe in the years following collapse. In concert with multi-wavelength ground- and space-based observations, numerous studies have utilized JWST data to constrain the dust formation around some of the nearest SNe \citep{Arendt2023, Hosseinzadeh23, Jones2023, Shahbandeh23, Bouchet2024, Gomez2024,  Matsuura2024,  Shahbandeh2024_22acko, Shahbandeh24, Zsiros24, Clayton2025, Sarangi2025, Szalai2025, Tinyanont2025}. 
This growing dataset is vital to understanding the onset of dust production in CCSNe and its impact on the evolution of the early Universe. 

In this paper, we present an analysis of newly acquired and archival multi-wavelength observations of SN~2017eaw, the first SN with multi-epoch late-time mid-infrared JWST observations. We review the observational data and reduction techniques in Section \ref{sec:obs}. In Section \ref{sec:analysis}, we analyze the UV and infrared photometry, determine the extent of mid-infrared evolution, and model the dust SED. We discuss the possible origin of SN~2017eaw's dust in Section \ref{sec:discussion} and conclude in Section \ref{sec:summary}. 

\section{Observations}\label{sec:obs}

\begin{figure*}
    \centering
    \includegraphics[width=\hsize]{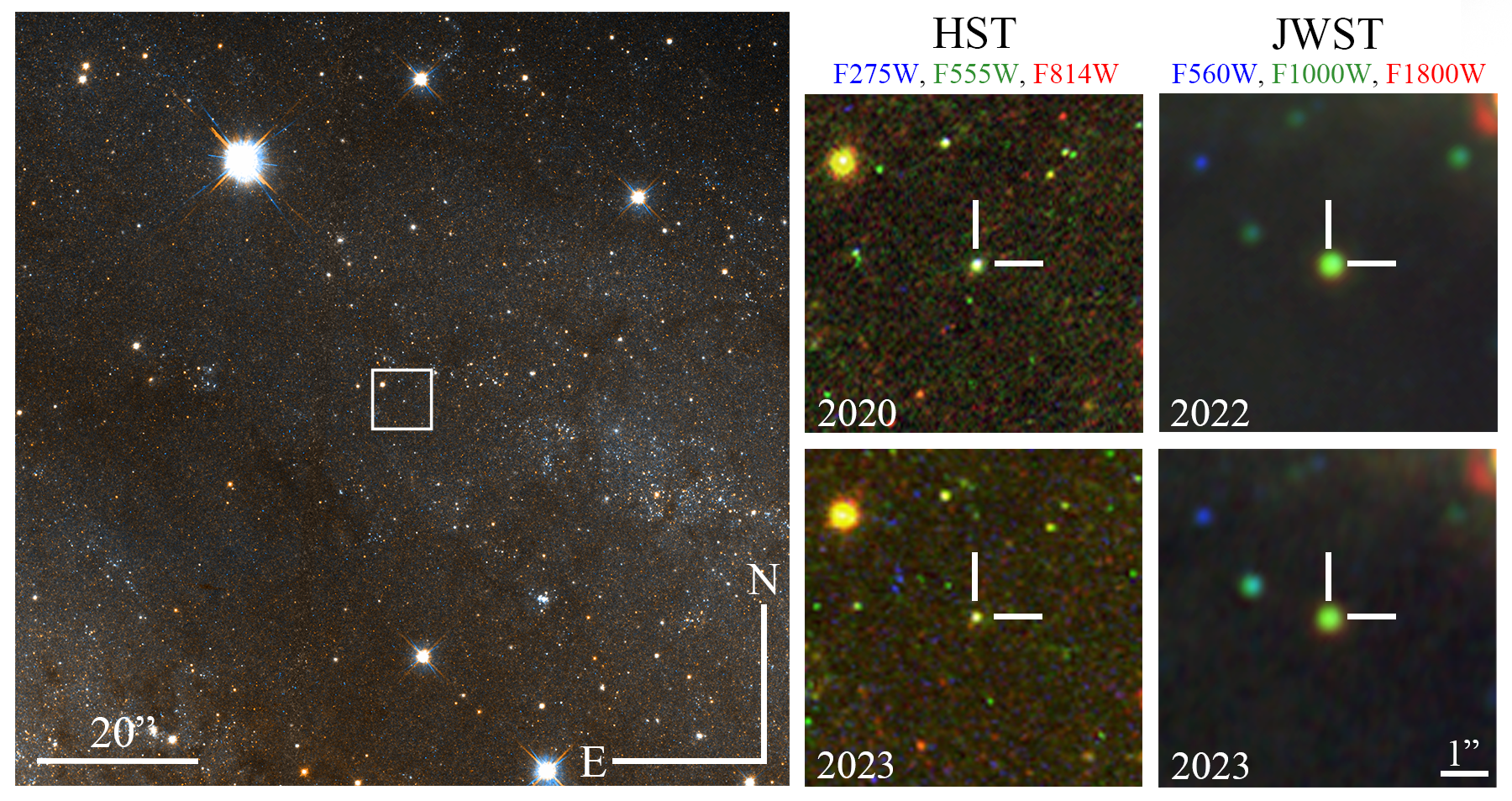}
    \caption{RGB false color images of SN~2017eaw using archival and recently acquired HST (F275W, F555W, and F814W) and JWST (F560W, F1000W, and F1800W) observations. The larger finder on the left was made with the F555W (60299 MJD) and F814W (59924 MJD) HST images taken in 2023.}
    \label{fig:cutouts}
\end{figure*}

SN~2017eaw was discovered on 2017 May 14 in NGC~6946, a nearby galaxy ($\mathrm{D}\approx7.12$ Mpc according to the latest TRBG distance\footnote{\url{https://edd.ifa.hawaii.edu/get_cmd.php?pgc=65001}, this distance is also used in \citet{Shahbandeh23}}) with a high SN rate (see Figure \ref{fig:cutouts}). SN~2017eaw has extensive multi-wavelength observations, both pre- and post-explosion \citep[explosion on 57885.2 MJD][]{VanDyk19}, and has been the subject of numerous studies \citep[e.g.][]{Tsvetkov18,Rho18,Kilpatrick18,Tinyanont19,VanDyk19,Szalai19,Rui19,Buta19,Weil20,Shahbandeh23,Bostroem23}. 

There is significant evidence of ejecta-CSM interaction in SN~2017eaw. Pre-explosion Spitzer Space Telescope (hereafter Spitzer) images show the progenitor star was surrounded by a dusty shell at $\sim$4000 $\mathrm{R}_\odot$ \citep{Kilpatrick18}.
Early time detections of SN~2017eaw in the X-ray, UV, and radio indicate moderate interaction between the SN shock and the CSM \citep{Szalai19}. Further, the optical light curve of SN~2017eaw exhibits a bump peaking around a week post-explosion that further suggests early time CSM interaction \citep{Szalai19}. Years after explosion there remain signs of ongoing ejecta-CSM interaction. Hubble Space Telescope (HST) near-UV imaging reveals the SN is still UV bright \citep[as shown in Figure \ref{fig:cutouts};][]{VanDyk23}{, light curves of the SN out to 1500 days indicate a shallower decline than expected for pure radioactive decay \citep{RizzoSmith2023},} and late-time ($>900$ days post-explosion) optical spectra exhibit boxy line profiles indicating that the ejecta is continuing to collide with the surrounding material \citep{Weil20, Shahbandeh23}. Given the continued detection of CSM interaction, it is likely that some pre-existing dust surrounds SN~2017eaw. 

In addition to the presence of dust in the CSM, there are several observational indicators that SN~2017eaw is producing dust in its ejecta. CO was detected in the near-infrared (NIR) spectra roughly one year post-explosion, demonstrating that the temperature of the ejecta has cooled enough for dust formation \citep{Rho18, Tinyanont19}. Further, nebular spectra of SN~2017eaw reveal blueshifted and asymmetric line profiles indicative of dust in the ejecta \citep{Rho18, Weil20, Shahbandeh23}. The likely presence of both pre-existing and newly-formed dust in SN~2017eaw makes it an ideal test case for understanding when newly-formed dust begins to dominate the infrared dust spectral energy distribution (SED). 

Post-explosion infrared observations of SN~2017eaw were executed by both Spitzer and JWST. Ground-based NIR and Spitzer observations (3.6 and 4.5 $\mu$m) at 200 and 500 days post-explosion and JWST mid-infrared observations at $\sim2000$ days reveal a population of observed silicate dust that has increased in mass over the years \citep[from $\sim1\times10^{-4}$ to $5.5\times10^{-4}$ M$_\odot$][]{Tinyanont19, Shahbandeh23}. However, given the proximity of the Spitzer observations to explosion and the limited wavelength coverage, it is difficult to determine if the Spitzer and JWST dust populations are related.

We present the second epoch of mid-infrared JWST imaging of SN~2017eaw, taken one year after the first, and compare it directly to the previously published epoch in order to determine the nature and origin of the dust.
To further understand the dust evolution of SN~2017eaw, we also collect and analyze additional new and archival data, including optical and near-ultraviolet HST observations and ground-based optical and infrared imaging and spectroscopy. 

\subsection{JWST/MIRI}\label{sec:jwstphot}

Initial JWST Mid-Infrared Instrument \citep[MIRI;][]{Bouchet2015, Rieke2015, Wright2023} imaging of SN~2017eaw taken using the F560W, F1000W, F1130W, F1280W, F1500W, F1800W, F2100W, and F2550W filters was obtained in September 2022 (1957.7 days post-explosion) as part of the Cycle 1 General Observers (GO) 2666 Program \citep{Cycle1JWST}. Photometry and analysis of this epoch was previously published in \citet{Shahbandeh23}. 

Further JWST/MIRI observations of SN~2017eaw were also obtained on 26 September 2023 (2328.2 days post-explosion) as part of the Cycle 2 GO 3295 Program \citep{Cycle2JWST}. These observations were taken with the complete set of MIRI filters (F560W, F770W, F1000W, F1130W, F1280W, F1500W, F1800W, F2100W, and F2550W), using the FULL array with a FASTR1 readout pattern, a 4-point dither pattern, and an exposure time of 111 seconds for all filters. 

Given that one focus of this work is on the flux evolution between the Cycle 1 and 2 observations, we opt to reanalyze photometry of the Cycle 1 data so that the methodology remains consistent between epochs. In this work, both the Cycle 1 and Cycle 2 JWST observations of SN~2017eaw were processed with the JWST Calibration Pipeline version 1.15.1, with the Calibration Reference Data System version 11.17.25 \citep{jwstpipe}. 

We attempt aperture photometry on the Cycle 1 and 2 images using several different methods, outlined in Appendix \ref{sec:ap_v_psf}. However, \citet{Shahbandeh23} report PSF photometry for SN~2017eaw, and we find that our aperture photometry methods result in flux values for the Cycle 1 observations which are 10-40\% higher than those reported in \citet{Shahbandeh23}.  
For consistency with the published photometry, we instead report PSF photometry done using \texttt{space\textunderscore phot}\footnote{\texttt{space\textunderscore phot} version 0.2.5 \url{https://space-phot.readthedocs.io}} \citep{spacephot, Pierel24}. We note that this choice does not significantly impact the conclusions of this work since they are based primarily on the difference between the JWST/MIRI epochs and not the absolute flux calibration. PSF photometry with \texttt{space\textunderscore phot} is done on the stage 2 products for all filters except F2550W. This involves fitting the SN's PSF in each of the 4 individual Level 2 CAL files using WebbPSF \citep[version 1.2.1]{WebbPSF1, WebbPSF2} models. Given the low signal to noise detection of SN2017eaw in F2550W, we opt to do PSF photometry on the Level 3 stacked images for this filter. The \texttt{space\textunderscore phot} routine for Level 3 photometry uses temporally and spatially dependent Level 2 PSF models from WebbPSF, and drizzles them together to create a Level 3 PSF model. 
While the MIRI PSF models have been significantly updated since the publication of \citet{Shahbandeh23}, we find our Cycle 1 PSF photometry is mostly consistent with the previously reported values (see Appendix \ref{sec:ap_v_psf} for more information).
We report the flux values from \texttt{space\textunderscore phot} for both the Cycle 1 and Cycle 2 observations in Table \ref{tab:17eawpsfphot}. 

\begin{table*}
 \caption{JWST/MIRI Cycle 1 and 2 observations of SN~2017eaw} \label{tab:17eawpsfphot}
 \hspace*{-1.6cm}
 \begin{tabular}{ l c c c c c c c c}
    \hline
     & \multicolumn{4}{c}{\bf{Cycle 1}} & \multicolumn{4}{c}{\bf{Cycle 2}} \\
     Filter & MJD & Phase\footnote{From explosion on 57885.2 MJD \citep{VanDyk19}} & Flux {Density} & AB Mag & MJD & Phase & Flux {Density} & AB Mag \\
     & & [days]& [{$\mathrm{\mu}$Jy}] & & & [days]& [{$\mathrm{\mu}$Jy}] & \\
     \hline
     F560W & 59842.885 & 1957.7 & {$5.99\pm0.14$} & $21.957\pm0.025$ &  60213.391 & 2328.2 & {$3.45\pm0.12$} & $22.555\pm0.038$ \\
    F770W & -- & -- & -- & -- & 60213.394 & 2328.2 & {$7.64\pm0.17$} & $21.692\pm0.024$ \\
    F1000W & 59842.896 & 1957.7 & {$54.79\pm0.15$} & $19.553\pm0.003$ &  60213.405 & 2328.2 & {$41.53\pm0.45$} & $19.854\pm0.012$ \\
    F1130W & 59842.906 & 1957.7 & {$53.63\pm0.62$} & $19.576\pm0.012$ &  60213.409 & 2328.2 & {$38.70\pm1.27$} & $19.931\pm0.035$ \\
    F1280W & 59842.913 & 1957.7 & {$42.76\pm0.66$} & $19.822\pm0.017$ &  60213.417 & 2328.2 & {$34.58\pm1.02$} & $20.053\pm0.032$ \\
    F1500W & 59842.919 & 1957.7 & {$60.42\pm1.14$} & $19.447\pm0.020$ &  60213.422 & 2328.2 & {$47.30\pm1.97$} & $19.713\pm0.044$ \\
    F1800W & 59842.927 & 1957.7 & {$101.49\pm3.48$} & $18.884\pm0.037$ &  60213.428 & 2328.2 & {$108.20\pm6.20$} & $18.814\pm0.060$ \\
    F2100W & 59842.935 & 1957.7 & {$112.15\pm6.69$} & $18.776\pm0.063$ &  60213.436 & 2328.2 & {$107.88\pm14.86$} & $18.818\pm0.140$ \\
    F2550W & 59842.941 & 1957.7 & {$106.67\pm10.24$} & $18.830\pm0.100$ &  60213.439 & 2328.2 & {$69.71\pm20.34$} & $19.292\pm0.278$ \\
    \hline
 \end{tabular}
\end{table*}

\subsection{HST Optical and UV}
 
Several HST observations of SN~2017eaw have been taken since explosion (see Table \ref{tab:17eawHSTphot}). \citet{VanDyk23} reported HST photometry of SN~2017eaw from late 2020 and early 2022 as part of their study of the progenitor. Since then observations in ACS/WFC F555W and F814W (PI: C. Kilpatrick, ID: 17070) and WFC3/UVIS F275W and F555W (PI: W. Jacobson-Galan, ID: 17506) in late 2022 and late 2023 respectively, have been completed. 

We use \texttt{DOLPHOT} \citep{Dolphin00, Dolphin16} to obtain PSF photometry of SN~2017eaw in all HST images. We use the calibrated and charge-transfer-efficiency (CTE) corrected \texttt{flc} and the corresponding drizzled \texttt{drc} frames from the Mikulski Archive for Space Telescopes (MAST) as inputs for \texttt{DOLPHOT}. Each epoch and filter combination was run through \texttt{DOLPHOT} separately and the \texttt{flc} frames were aligned to the associated \texttt{drc} image. We use the same \texttt{DOLPHOT} parameter settings as were used for the HST PHAT survey \citep{Dalcanton12, Williams14}. \texttt{DOLPHOT} detected a ``good" star (``object type"=1) at the location of SN~2017eaw in all filters and epochs. Where available, we find our photometry is completely consistent with published values in \citet{VanDyk23}. 
We present the PSF photometry of the detected source in Table \ref{tab:17eawHSTphot} for SN~2017eaw. All UV, optical, and NIR magnitudes are reported in Vega magnitudes. 

\begin{table}
\centering
 \caption{Late time optical and NIR imaging observations of SN~2017eaw} \label{tab:17eawHSTphot}
\hspace*{-1.6cm}
 \begin{tabular}{ c c c c c}
    \hline
    Filter & MJD & Phase & Vega Mag & Tele/Inst\\
    & & [days] & & \\
    \hline
    F336W$^*$ & 59156.290 & 1271.1 &  $24.28\pm0.05$ & HST/WFC3\\
    F275W$^*$ & 59156.389 & 1271.2 & $22.75\pm0.02$ & HST/WFC3\\
    F555W$^*$ & 59164.831 & 1279.6 & $23.77\pm0.02$ & HST/WFC3\\
    F814W$^*$ & 59164.813 & 1279.6 & $23.12\pm0.03$ & HST/WFC3\\
    F555W$^*$ & 59622.265 & 1737.1 & $23.96\pm0.02$ & HST/WFC3\\
    F814W$^*$ & 59622.259 & 1737.1 & $23.30\pm0.03$ & HST/WFC3\\
    F555W & 59924.463 & 2039.3 & $24.02\pm0.03$ & HST/ACS\\
    F814W & 59924.457 & 2039.3 & $23.35\pm0.02$ & HST/ACS\\
    F275W & 60299.132 & 2413.9 & $24.75\pm0.13$ & HST/WFC3\\
    F555W & 60299.141 & 2413.9 & $24.20\pm0.02$ & HST/WFC3\\
    K & 60040.505 & 2155.3 & $19.53\pm0.10$ & MMT/MMIRS\\
    J & 60044.427 & 2159.2 & $21.73\pm0.09$ & MMT/MMIRS\\
    H & 60044.488 & 2159.3 & $>19.96$ & MMT/MMIRS\\
    J & 60282.062 & 2396.9 & $21.59\pm0.19$ & MMT/MMIRS\\
    K & 60282.092 & 2396.9 & $>19.23$ & MMT/MMIRS\\
    r$^\dagger$ & 60203.144 & 2317.9 & $22.95\pm0.06$ & MMT/Binospec\\
    i$^\dagger$ & 60205.257 & 2320.1 & $23.00\pm0.10$ & MMT/Binospec\\
    \hline
 \end{tabular}
 \begin{tablenotes}
      \small
      \item $^*$ Previously published in \citet{VanDyk23}
      \item $^\dagger$ AB magnitudes are r: $23.11\pm0.06$ and i: $23.37\pm0.10$
\end{tablenotes}
\end{table}

\subsection{MMT Optical and NIR Imaging}
Additionally, we report optical and NIR ground-based photometry of SN~2017eaw. We present \textit{r} and \textit{i} band imaging of SN~2017eaw from 16 and 18 September 2023 (60203.144 and 60205.257 MJD) respectively, taken with the Binospec instrument on the MMT \citep{Binospec}, and NIR \textit{JHK} photometric observations of SN~2017eaw taken with the MMT and Magellan Infrared Spectrograph (MMIRS) on the MMT \citep{MMIRS} in Spring 2023 and December 2023.  

For Binospec observations, we utilize a standard dither pattern. These images are then reduced using a custom python Binospec imaging reduction pipeline\footnote{Initially written by K. Paterson and available on GitHub: \url{https://github.com/CIERA-Transients/POTPyRI}}, which does standard flat-fielding, sky background estimation, astrometric alignments, and stacking of the final individual exposures. 

For MMIRS, each observation consisted of a dithered sequence which alternates between the target field and a off-galaxy field to allow for better sky subtraction given the IR-brightness of NGC6946. The resulting $J$, $H$, and $K$ band observations were reduced using a custom pipeline\footnote{Adapted from the MMIRS imaging pipeline available on github: \url{https://github.com/CIERA-Transients/POTPyRI}} which does standard dark-current correction, flat-fielding, sky background estimation and subtraction, astrometric alignments, and stacking of the final individual exposures. 

The total field-of-view (FOV) of MMIRS (6'.9$\times$6'.9) and Binospec (two 8'$\times$15' FOVs with 3' gap) is large enough to calibrate photometric zeropoints using isolated stars with cataloged Two Micron All Sky Survey \citep[2MASS;][]{2MASS} and Panoramic Survey Telescope and Rapid Response System \citep[Pan-STARRS;][]{PanSTARRS}, respectively. We derive an effective (e)PSF model for each image by fitting bright, isolated stars with the \texttt{EPSFBuilder} tool from the \texttt{photutils} package in \texttt{Astropy}. For all filters where the supernova is detected, we then perform PSF-fitting at the location of the target as well as a set of 20 or more stars spread throughout the image. A low-order, two-dimensional polynomial is included in the fit to account for any spatially varying background and avoid over-fitting of the stars. To estimate the statistical uncertainty of each flux measurement, we first set the statistical uncertainty per pixel using the RMS error of the fit residuals scaled by a factor of the square root of the reduced $\chi^2$ (usually $\gtrsim$1), then multiply by the number of `noise pixels' of the ePSF\footnote{A derivation of this quantity by F. Masci can be found here: \url{http://web.ipac.caltech.edu/staff/fmasci/home/mystats/noisepix_specs.pdf}}. We use the set of 2MASS or Pan-STARRS calibration stars to derive aperture corrections ($\lesssim$0.1 mag in all filters) to scale PSF-fitting magnitudes to the images' photometric zeropoints. We adopt the statistical flux uncertainty summed in quadrature with the RMS error of the stars used in the zeropoint and ePSF aperture correction as the total uncertainty in our reported magnitudes. Despite the large FOV of both MMIRS and Binospec, the limited number of isolated 2MASS and Pan-STARRS stars means that the zeropoint RMS dominates the reported error. 

SN~2017eaw was not detected in the MMIRS \textit{H}-band and \textit{K}-band images taken on 60044.427 and 60282.092 MJD (2159 and 2397 days post-explosion), respectively. For these observations, we instead report a 5$\sigma$ limiting magnitude, based off randomly placed background apertures near the location of SN~2017eaw. 
Optical (converted to Vega magnitudes) and NIR MMT photometry of SN~2017eaw are reported in Table \ref{tab:17eawHSTphot}.

\subsection{Keck LRIS Spectroscopy}
To complement the Cycle 2 JWST MIRI observations, we obtained an optical spectrum of SN~2017eaw on 2024 Aug 31 (60553.25 MJD, 2668 days post-explosion) using the Low Resolution Imaging Spectrometer \citep[LRIS;][]{Oke95} on Keck I. The spectrum was taken with a 1.5" slit width with the 600/4000 grism and the 400/8500 grating at a central wavelength of 7700 \AA\ and a total exposure time of 7200 seconds. 

The spectrum was reduced in a standard way using \texttt{LPipe} \citep{Perley2019}. The complete spectrum is further discussed in Section \ref{sec:spectra}. 

\section{Analysis}\label{sec:analysis}

\begin{figure*}
    \centering
    \includegraphics[width=\hsize]{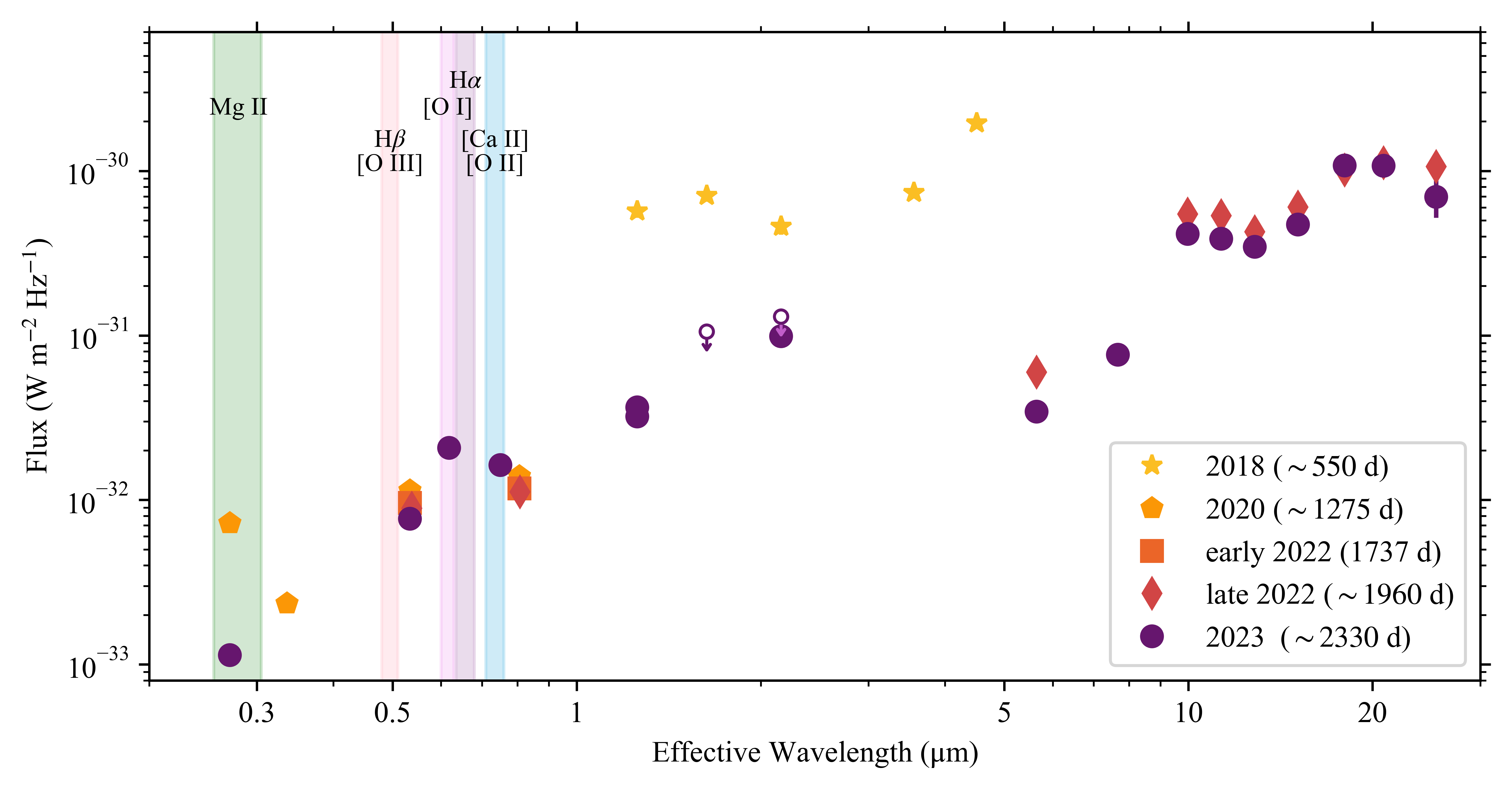}
    \caption{The evolution of the full SED of SN~2017eaw from 2020 to late 2023. The {2018 SED, which includes 3.6 and 4.5 \micron\ photometry from Spitzer and coincident \textit{JHK} photometry from the the Wide field InfraRed Camera at Palomar Observatory,} is also included for reference \citep{Tinyanont19}. Different markers indicate the different epochs, these epochs are defined to include observations taken $<$6 months from each other, i.e. ``late 2022" denotes the second half of the year. Note that the $r$ and $i$ band photometry (purple points at 0.62 and 0.75 \micron) are elevated due to the presence of broad nebular lines, most notably H$\alpha$, in the filter bandpasses. The flux across all bands $<$18 \micron\ has decreased in each consecutive epoch. }
    \label{fig:fullSED}
\end{figure*}

We present the full time series spectral energy distribution (SED) of SN~2017eaw from 2020 to late 2023 in Figure \ref{fig:fullSED}. The flux has generally decreased in each consecutive epoch. The notable exception to this trend are the $r$ and $i$ bands, where the filter bandpasses include broad hydrogen, calcium, and oxygen lines which have been observed in nebular spectra of SN~2017eaw \citep[see Section \ref{sec:spectra} and][]{Shahbandeh23}. Strikingly, the mid-infrared fluxes blue-ward of 18 $\mu$m have declined between Cycle 1 and Cycle 2 but are consistent between Cycle 1 and 2 for wavelengths $\geq18$ $\mu$m. As we discuss in Section \ref{sec:compstars} below, it is difficult to determine the statistical significance of the SED evolution redward of 15 $\mu$m. 

\subsection{Comparison with Stars in the Field} \label{sec:compstars}
To ensure that the observed decrease in luminosity is the result of a true decrease in flux and not the result of changes in the different observing parameters (i.e. exposure time, camera orientation, etc.) used in the Cycle 1 and Cycle 2 observations, we perform PSF photometry on several stars in the field using methods similar to those used for the SN~2017eaw photometry reported in Table \ref{tab:17eawpsfphot}. 

We first identify {infrared} bright objects with minimal variability in the field by using the 2MASS catalog \citep{2MASS}. These objects are then confirmed to be point sources in all of the JWST filters. Since star clusters may appear as point sources in MIRI filters, we also include a cut to remove any objects in crowded regions by using an HST F814W image of NGC6946 as a reference. This procedure results in 9 comparison stars for the SN~2017eaw field. Given the low signal to noise detections of these reference stars, particularly in the redder bands, we opt to do photometry on the stage 3 products for all filters rather than the stage 2 products as was done for SN~2017eaw. We note that stage 2 and 3 photometry of SN~2017eaw produce flux change measurements that are consistent within the uncertainties. 

The percent difference in flux between the Cycle 1 ($C1$) and Cycle 2 ($C2$) comparison star observations is calculated as 
$(C2 - C1)/C1$. To determine percent change in the total comparison sample we calculate the average change across the sample and adopt the standard deviation of the flux values as the error in this measurement. 

\begin{figure}
    \centering
    \includegraphics[width=\hsize]{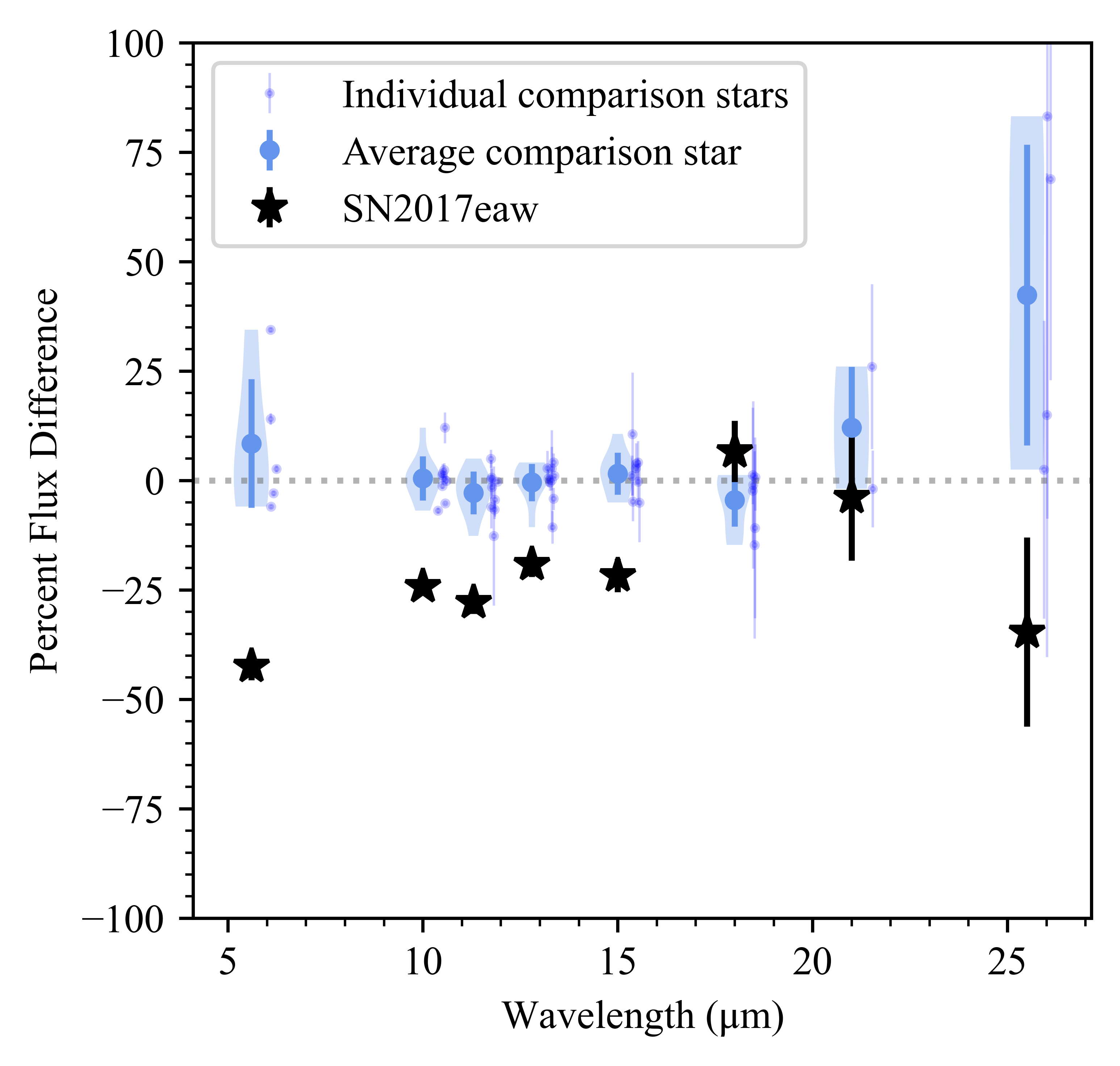}
    \caption{Percent change in flux from Cycle 1 to Cycle 2 for comparison stars in the FOV of SN~2017eaw. The width of the violin plot denotes the number density of comparison stars at a given flux difference, error bars denote one sigma from the average change in flux. For reference, the individual reference stars (small points) are offset $\sim$0.5 \micron\ from the average (large points). The stars in the field are consistent with zero change in flux between epochs for all filters except 25 $\mu$m, though the reddest bands have significant scatter and smaller sample sizes.}
    \label{fig:compstars}
\end{figure}

As shown in Figure \ref{fig:compstars}, we find that the comparison stars are consistent with no change in flux across the two epochs in the F560W, F770W, F1000W, F1130W, F1280W, and F1500W filters. The change in flux is consistent with zero redward of 15 $\mu$m as well, with the exception of the F2550W filter. Filters where fewer reference stars are detected have larger errors due to small number statistics. For example, one of the reference stars has a high variance in F560W resulting in large error bars on the average flux change for this filter. Due to the high sky flux in the redder bands ($\geq$18 $\mu$m) there are significant errors in flux measurements for individual stars, large scatter between stars, and smaller sample sizes as few of our reference stars are detectable in the reddest bands.
Therefore we can not make a high significance measurement of the extent of flux change in SN~2017eaw redward of 15 $\mu$m.

\subsection{Decrease in UV Flux}
HST F275W and F336W observations from 2020 indicated that SN~2017eaw was UV bright. \citet{VanDyk23} suggested the elevated UV flux was the result of CSM-ejecta interaction but could not rule out the possibility of an underlying UV source like an O-star or small stellar cluster. 
As shown in Figure \ref{fig:fullSED}, the F275W observations taken in late 2023 reveal that, while the UV is still elevated, it has declined significantly.
The F275W flux in 2023 is 16\% of the F275W flux in 2020. Such a significant UV evolution is unlikely to be caused by an underlying star cluster or main sequence star. 

It is notable that the F275W filter is particularly sensitive to CSM-ejecta interaction given it includes the Mg II $\lambda\lambda$2796, 2803 doublet \citep{Dessart23}. Boxy H$\alpha$ emission lines, a signature of late time CSM interaction, have been observed in the optical spectra of SN~2017eaw since 900 days post-explosion \citep{Weil20}, well before the 2020 F275W observations, and boxy H$\alpha$ is still present in more recent spectra (see Section \ref{sec:spectra}). Given the evolution and additional observational signatures, the UV evolution is very likely tracing the CSM-ejecta interaction. 

\subsection{Progenitor Disappearance}

\begin{figure}
    \centering
    \includegraphics[width=\hsize]{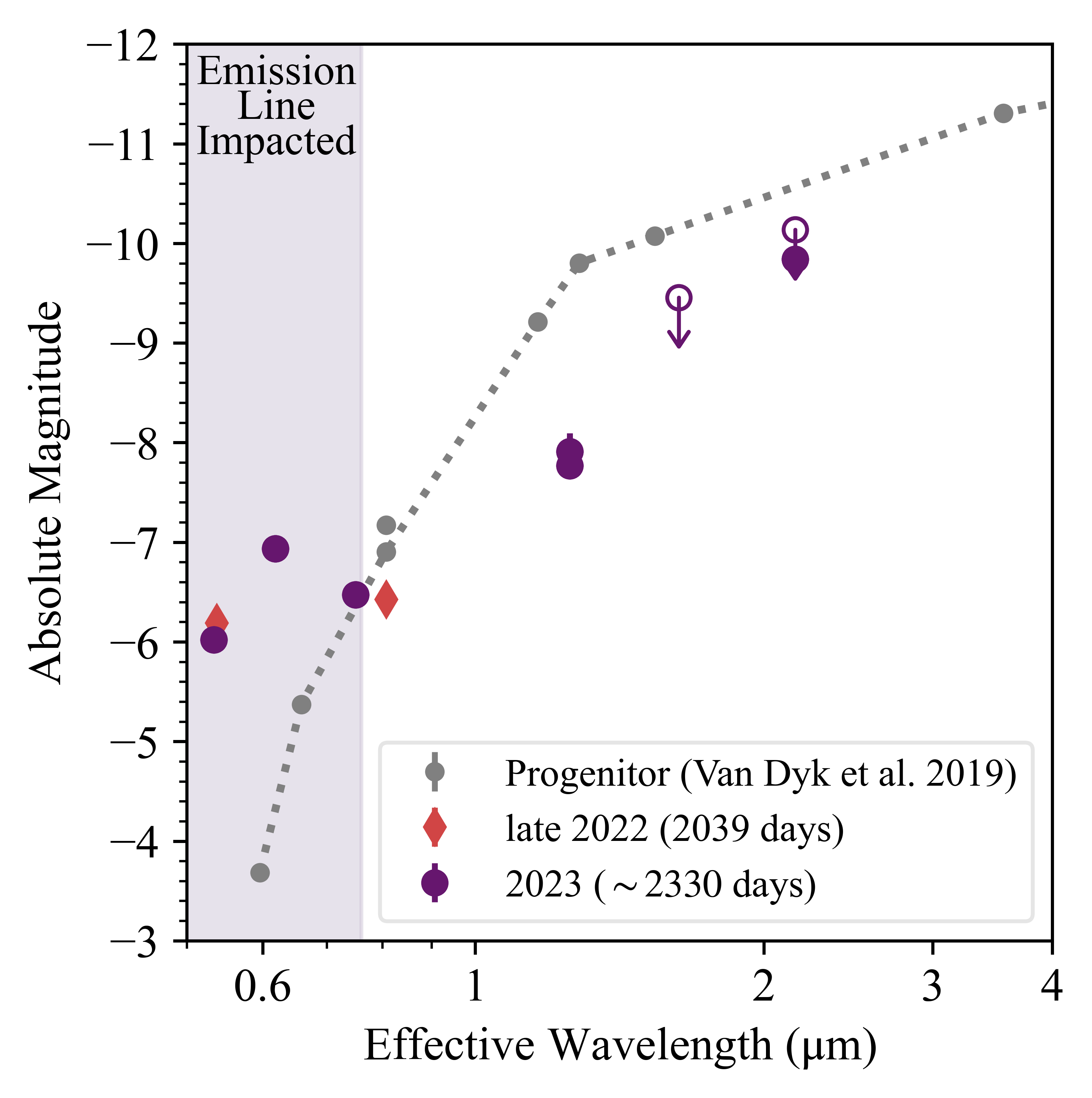}
    \caption{Comparison of optical and NIR observations of SN~2017eaw from 2023 and late 2022 with the progenitor photometry reported in \citet{VanDyk19}. $J$ and $K$ -band (and F814W to some degree) observations indicate that the SN has faded significantly below the progenitor level.}
    \label{fig:prodispearence}
\end{figure}

Given the proximity of the host galaxy, numerous images of the progenitor of SN~2017eaw were taken by both ground and space-based observatories. HST and Spitzer images identified the progenitor as a 12-15 M$_\odot$ dusty red supergiant \citep{Kilpatrick18, VanDyk19, Rui19}. Since the supernova explosion, several epochs of HST photometry have been acquired. An analysis of the post-explosion imaging up to February 2022, suggested that the F814W flux had faded below the progenitor level \citep{VanDyk23}. 

A further epoch of F814W imaging was obtained in December 2022 and we find it is similarly below the progenitor flux. Additionally, we compare the NIR progenitor flux to the MMT/MMIRS photometry from December 2023.
As shown in Figure \ref{fig:prodispearence}, we find that the $J$ and $K$ band detections are $>$2 magnitudes fainter than the progenitor. We are therefore able to confirm the progenitor identification and verify that it has significantly faded from pre-supernova observations.

\subsection{Dust Modeling Methods}\label{sec:dustmod}

We assume the mid-infrared flux observed by JWST/MIRI in Cycle 1 and Cycle 2 is due to thermal emission from dust grains in or near the SN ejecta. To model the cool and hot dust components we employ several analytical dust models using a procedure adapted from \citet{Hosseinzadeh23}.

Given the prominent shape of the silicate feature at $\sim10$ $\mu$m, the observed dust is unlikely to be optically thick. Therefore we first fit the dust using an optically thin model as was done in \citet{Shahbandeh23} and \citet{Zsiros24}. We also model a dusty sphere \citep[similar to ][]{Shahbandeh23} and a dusty shell motivated by work presented in \citet{Dwek2024}; for both of these dust models we allow the optical depth to vary. All of the models presented here assume there are two temperature components, as motivated by the SED shape, within the same geometry. Details of the luminosity equations for these three models can be found in Appendix \ref{sec:dusteqs}.

We fit all three models (Equations \ref{eq:L_thin}, \ref{eq:L_sphere}, and \ref{eq:L_shell}) to a filter-integrated model of the observed SED using the MCMC routine implemented in the Light Curve Fitting package \citep{lightcurvefitting2}. Additionally, we fit an intrinsic scatter term, $\sigma$, which inflates the error bars on each photometric data point by a factor of $\sqrt{1+\sigma^2}$, to account for underestimated photometric uncertainties, and to account for uncertainties in the model (e.g. infrared line emission). We run 20 walkers for 2000 steps to reach convergence and then 1000 more steps to properly sample the posterior. All of the optical filters blueward of 0.8 $\mu$m include flux from broad emission lines (see Figure \ref{fig:fullSED}), so we exclude all of these filters when fitting the SED. We include only the NIR detections and the JWST/MIRI observations in our fit of the Cycle 2 (2023) SED. For Cycle 1, only the JWST/MIRI observations are considered. 

As shown in Figure \ref{fig:fullSED}, the 5-25 $\mu$m SED exhibits two distinct peaks with a trough at $\sim$13 $\mu$m. This double humped shape is characteristic of optically thin silicate dust. Therefore we assume that the cool component is silicate dust, with $\rho_\mathrm{silicate} = 3.3$ g cm$^{-3}$ and $a= 0.1\ \mu$m as given by \citet{Laor93}. The composition of the hot component is not as clearly identified. Recent studies report both amorphous carbon and silicate hot dust components around CCSNe \citep{Hosseinzadeh23, Shahbandeh23, Zsiros24}, so we try models with both silicate and amorphous carbon dust {($a= 0.1\ \mu$m and $\kappa_\nu$ from \citealt{Draine1985} and \citealt{Colangeli95} for silicate and amorphous carbon respectively)} for the optically thin dust case. For the dusty sphere and dusty shell cases, we assume both components are silicate dust. 

\begin{figure}
    \centering
    \includegraphics[width=\hsize]{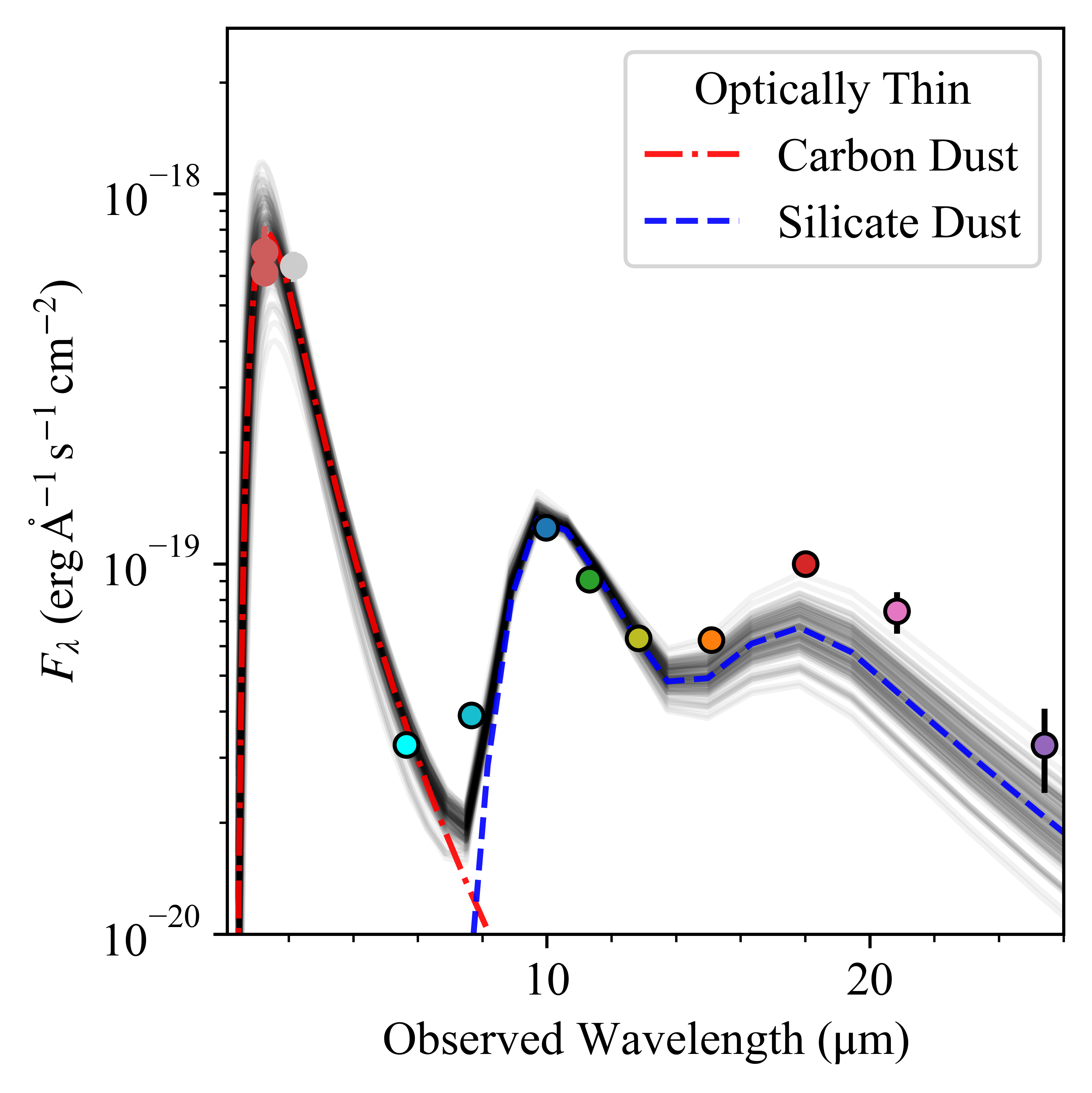}
    \caption{The best-fitting hot amorphous carbon and cool silicate optically thin dust model for the full Cycle 2 {infrared} SED. Colored points indicate different filters. The carbon+silicate model can not reproduce the observed flux in the reddest wavelengths and requires temperatures (T$=2200\pm200 $ K) significantly hotter than the dust condensation temperature to fit the hot dust component. }
    \label{fig:CSiModel}
\end{figure}

We attempt to fit a model of hot amorphous carbon dust and cool silicate optically thin dust, as was done for SN~1980K \citep{Zsiros24}, to the NIR and JWST/MIRI photometry from Cycle 2. 
As shown in Figure \ref{fig:CSiModel}, we find that the carbon+silicate model can not reproduce the photometry $>$15 $\mu$m. We note that this excess could potentially be fit with an additional cooler component, as might be expected of dust formed in the ejecta or pre-existing dust at larger radii. However, given that the SED evolution redward of 15\micron\ is not well constrained, we avoid implementing a third dust component. 
The hot carbon component requires a temperature of T$_\mathrm{hot}=2200\pm200 $ K to reproduce the observed NIR peak, this is significantly above the condensation temperature of amorphous carbon dust \citep[T$ \approx1500$ K;][]{Lodders1997}. We therefore deem it unlikely that the hot dust component in SN~2017eaw is primarily carbonaceous dust. However, we cannot rule out the possibility that there is both carbon and silicate dust in the ejecta, though the shape and temperature of the SED would suggest a high ratio of silicate to carbon dust if carbon dust is present. 

\begin{figure*}
    \centering
    \includegraphics[width=\hsize]{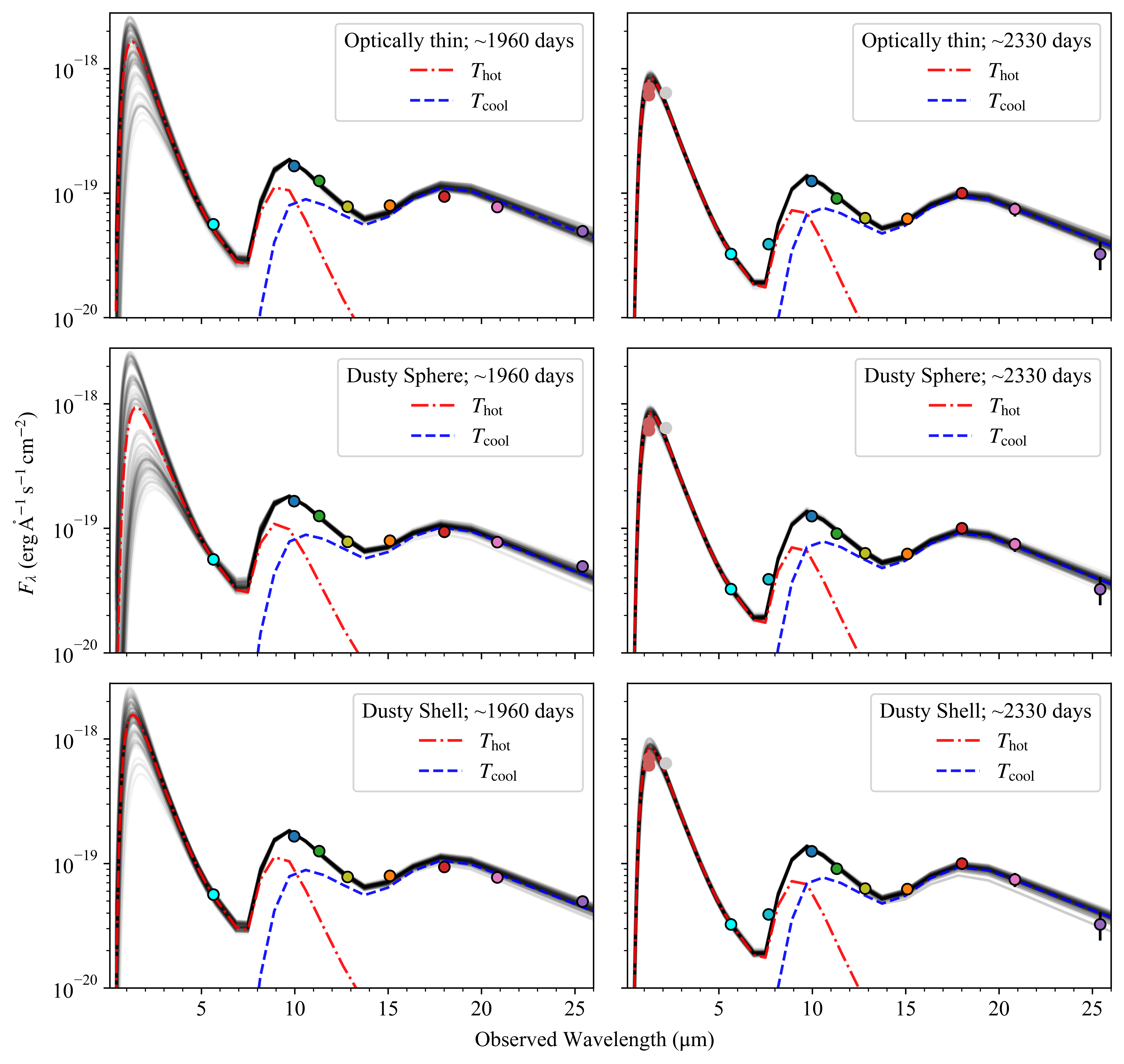}
    \caption{The best-fitting double silicate dust models for the full Cycle 1 and Cycle 2 {infrared} SEDs for the optically thin, dusty sphere, and dusty shell models. All three geometries are able to reproduce the observed SED. Note that the addition of NIR photometry significantly constrains the hotter dust component of the Cycle 2 dust. A complete figure set of corner plots for these models are available (see Figure \ref{fig:corner}).}
    \label{fig:dustmodels}
\end{figure*}

\begin{figure}
\centering
\includegraphics[width=\hsize]{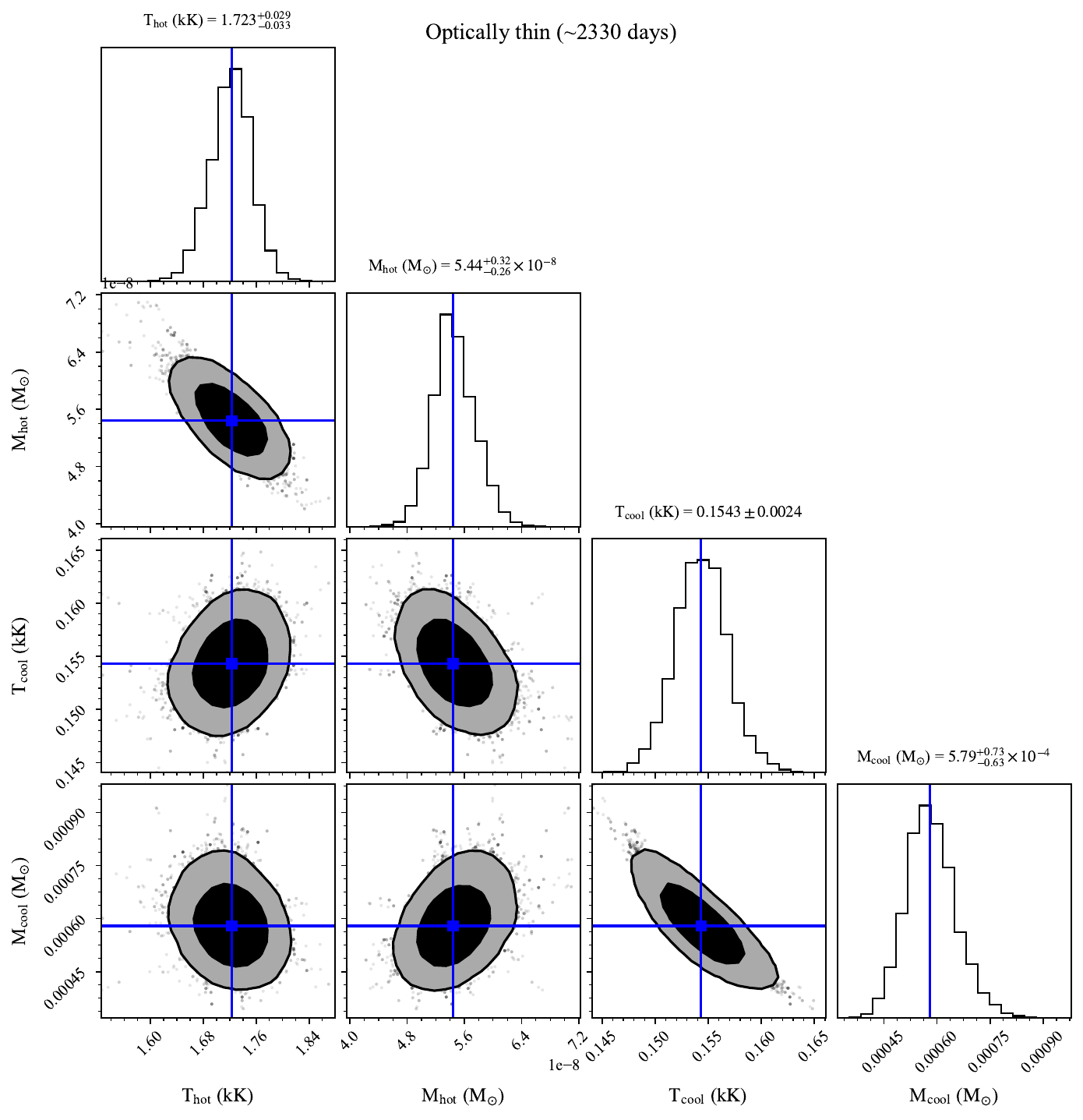}
\caption{The corner plot for the Cycle 2 optically thin dust model. All 6 corner plots corresponding to the SEDs displayed in Figure \ref{fig:dustmodels} are available as a figure set in the online journal.}
\label{fig:corner}
\end{figure}

Given the poorer quality fit of the carbon+silicate optically thin model, we focus primarily on models where both dust components are silicate. 
In Table \ref{tab:dustparameters}, we list the model parameters, their priors, and their best-fit values (median and 1$\sigma$ equal-tailed credibility interval), for the optically thin, dusty sphere, and dusty shell models of the full considered SED for Cycle 1 and Cycle 2. The best-fit models for all three iterations, and the separate dust components, are shown in Figure \ref{fig:dustmodels} (the corresponding corner plots are available in the online journal as a figure set accompanying Figure \ref{fig:corner}). Since the Cycle 1 SED does not include F770W or NIR observations, we also fit the Cycle 2 SED excluding these filters. For completeness, we also present the results of the Cycle 2 fits excluding the NIR and F770W observations in Table \ref{tab:dustparameters} and discuss them further in Appendix \ref{sec:no770NIR}.

\begin{table*}[t]
\centering
\caption{Model {parameters for dust with two silicate components}} \label{tab:dustparameters}
 \begin{threeparttable}
     \begin{tabularx}{0.95\textwidth}{ l l c c c c}
        \hline
         & & & \bf{Cycle 1 ($\sim$1960 days)} & \multicolumn{2}{c}{\bf{Cycle 2 ($\sim$2330 days)}} \\
         
         & Parameter & Priors & All Filters$^a$ & All Filters$^b$ & No NIR \& F770W$^a$\\
         \hline
         \parbox[t]{2mm}{\multirow{5}{*}{\rotatebox[origin=c]{90}{Optically Thin}}}& T$_\mathrm{hot}$ $[\mathrm{kK}]$& Uniform(1.0, 2.0) & $1.8^{+0.1}_{-0.2}$ & $1.72 \pm 0.03$ & $1.6 \pm 0.3\ $\\
        & T$_\mathrm{cool}$ $[\mathrm{kK}]$& Uniform(0.05, 0.3)&$ 0.155^{+0.004}_{-0.003}$ & $0.154 \pm 0.002$ & $0.153^{+0.004}_{-0.005}$\\
        & M$_\mathrm{hot}$ $[ \mathrm{M_{\odot}}]$& Log-Uniform(1e-15, 1.)&$ 7.6^{+1.9}_{-0.9} \times 10^{-8}\ $ & $5.4 \pm 0.3 \times 10^{-8}\ $ & $7^{+4}_{-2}\times 10^{-8}\ $ \\
        & M$_\mathrm{cool}$ $[ \mathrm{M_{\odot}}]$& Log-Uniform(1e-15, 1.)& $ 7 \pm 1 \times 10^{-4}\ $ & $ 5.8^{+0.7}_{-0.6} \times 10^{-4}\ $ & $6.1^{+1.6}_{-0.9}\times 10^{-4}\ $\\
        & Intrinsic scatter& Gaussian(0., 20.)& $2.7^{+0.6}_{-0.4}$ & $0.6^{+0.5}_{-0.4}$ & $0.4^{+0.5}_{-0.3}$\\
        \hline
         \parbox[t]{2mm}{\multirow{6}{*}{\rotatebox[origin=c]{90}{Dusty Sphere}}}& T$_\mathrm{hot}$ $[\mathrm{kK}]$& Uniform(0.5, 2.0) & $1.6^{+0.3}_{-0.4}$ & $1.73 \pm 0.04$ & $1.1 \pm 0.2\ $\\
        & T$_\mathrm{cool}$ $[\mathrm{kK}]$& Uniform(0.05, 0.3)&$ 0.163 \pm 0.005$ & $0.158^{+0.003}_{-0.002}$ & $0.151 \pm 0.005$\\
        & M$_\mathrm{hot}$ $[ \mathrm{M_{\odot}}]$& Log-Uniform(1e-15, 0.01)& $1.1^{+1.2}_{-0.3} \times 10^{-7}\ $ & $5.4^{+0.4}_{-0.3} \times 10^{-8}\ $ & $1.3^{+0.9}_{-0.5} \times 10^{-7}\ $\\
        & M$_\mathrm{cool}$ $[ \mathrm{M_{\odot}}]$& Log-Uniform(1e-15, 0.01)& $ 5.2^{+1.1}_{-0.9} \times 10^{-4}\ $ & $5.1 \pm 0.7 \times 10^{-4}\ $ & $7^{+2}_{-1} \times 10^{-4}\ $\\
        & R$_\mathrm{outer}$ $[10^{3}\ \mathrm{R_{\odot}}]$ & Log-Uniform(1, 2000.) & $ 700^{+200}_{-100}$ & $1500^{+300}_{-400}$ & $1300^{+500}_{-400}$\\
        & Intrinsic scatter& Gaussian(0., 20.)&$ 1.8^{+0.5}_{-0.4}$ & $0.8 \pm 0.5$ & $0.4^{+0.5}_{-0.3}$\\
        \hline
        \parbox[t]{2mm}{\multirow{7}{*}{\rotatebox[origin=c]{90}{Dusty Shell}}}& T$_\mathrm{hot}$ $[\mathrm{kK}]$& Uniform(1.0, 2.0) & $1.8 \pm 0.1$ & $1.73 \pm 0.03$ & $1.3 \pm 0.2$\\
        & T$_\mathrm{cool}$ $[\mathrm{kK}]$& Uniform(0.05, 0.3)&$ 0.158^{+0.003}_{-0.002}$ & $0.156^{+0.003}_{-0.002}$ & $0.151^{+0.003}_{-0.004}$\\
        & M$_\mathrm{hot}$ $[ \mathrm{M_{\odot}}]$& Log-Uniform(1e-10, 1.)& $9 \pm 1 \times 10^{-8}\ $ & $5.4 \pm 0.3 \times 10^{-8}\ $ & $9^{+3}_{-2} \times 10^{-8}\ $\\
        & M$_\mathrm{cool}$ $[ \mathrm{M_{\odot}}]$& Log-Uniform(1e-8, 1.)& $ 5.9^{+0.8}_{-0.5} \times 10^{-4}\ $ & $5.5 \pm 0.7 \times 10^{-4}\ $ & $7 \pm 1 \times 10^{-4}\ $\\
        & R$_\mathrm{inner}$ $[10^{3}\ \mathrm{R_{\odot}}]$& Log-Uniform(0, 2000.)   & $<1150$ & $<350$ & $<720$\\
        & R$_\mathrm{outer}$ $[10^{3}\ \mathrm{R_{\odot}}]$ & Log-Uniform(1, 5000.) & $ 1100^{+600}_{-200}$ & $2700^{+1300}_{-900}$ & $3000 \pm 1000$\\
        & Intrinsic scatter& Gaussian(0., 20.)&$ 2.1^{+0.6}_{-0.5}$ & $0.7^{+0.5}_{-0.4}$ & $0.5^{+0.5}_{-0.3}$\\
        \hline
    \end{tabularx}
    \begin{tablenotes}
      \small
      \item $^a$ includes same filters as Cycle 1 -- full JWST/MIRI filter suite excluding F770W
      \item $^b$ includes $J$ and $K$ in addition to full JWST/MIRI filter suite
    \end{tablenotes}
 \end{threeparttable}
\end{table*}

\noprint{\figsetstart}
\noprint{\figsetnum{7}}
\noprint{\figsettitle{Corner plots for the three model geometries for Cycles 1 and 2.}}

\figsetgrpstart
\figsetgrpnum{7.1}
\figsetgrptitle{Dusty Shell; ~1960 days}
\figsetplot{corner_plot_Dusty_Shell_1960_days.pdf}
\figsetgrpnote{Corner plot for Cycle 1 (~1960 days) Dusty Shell model.}
\figsetgrpend

\figsetgrpstart
\figsetgrpnum{7.2}
\figsetgrptitle{Dusty Shell; ~2330 days}
\figsetplot{corner_plot_Dusty_Shell_2330_days.pdf}
\figsetgrpnote{Corner plot for Cycle 2 (~2330 days) Dusty Shell model.}
\figsetgrpend

\figsetgrpstart
\figsetgrpnum{7.3}
\figsetgrptitle{Dusty Sphere; ~1960 days}
\figsetplot{corner_plot_Dusty_Sphere_1960_days.pdf}
\figsetgrpnote{Corner plot for Cycle 1 (~1960 days) Dusty Sphere model.}
\figsetgrpend

\figsetgrpstart
\figsetgrpnum{7.4}
\figsetgrptitle{Dusty Sphere; ~2330 days}
\figsetplot{corner_plot_Dusty_Sphere_2330_days.pdf}
\figsetgrpnote{Corner plot for Cycle 2 (~2330 days) Dusty Sphere model.}
\figsetgrpend

\figsetgrpstart
\figsetgrpnum{7.5}
\figsetgrptitle{Optically thin; ~1960 days}
\figsetplot{corner_plot_Optically_thin_1960_days.pdf}
\figsetgrpnote{Corner plot for Cycle 1 (~1960 days) Optically thin model.}
\figsetgrpend

\figsetgrpstart
\figsetgrpnum{7.6}
\figsetgrptitle{Optically thin; ~2330 days}
\figsetplot{corner_plot_Optically_thin_2330_days.pdf}
\figsetgrpnote{Corner plot for Cycle 2 (~2330 days) Optically thin model.}
\figsetgrpend

\figsetend

\subsection{Dust Modeling Results}

Two silicate dust components are able to reproduce the complete SED for all three model geometries, as shown in Figure \ref{fig:dustmodels} (top: optically thin, middle: dusty sphere, bottom: dusty shell). While the F770W observation somewhat reduces the spread in the posterior distribution and constrains the hot dust component, the NIR photometry is the most significant factor in constraining the dust models given that the hot dust SED peaks near the effective wavelength of the $J$ filter. The particular dust model is not a significant factor in constraining the dust properties, all the models agree on the masses ($\sim5.5\times10^{-4}\ \mathrm{M}_\odot$ for the cool and $\sim5.4\times10^{-8}\ \mathrm{M}_\odot$ for the hot dust) and temperatures ($\sim160$ K for the cool and $\sim1700$ K for the hot) of the components for either epoch. The inner radius in the dusty shell model is consistent with zero (i.e. a dusty sphere) for both Cycle 1 and 2 observations. Given this, we report the $3\sigma$ upper limit on R$_\mathrm{inner}$ in Table \ref{tab:dustparameters}. 

When we compare the Cycle 1 to the Cycle 2 full filter set models, we find that all of the best-fit values are consistent within error with the exception of the $\leq3\sigma$ differences in M$_\mathrm{hot}$ (mass of hot dust component) for all three model geometries and R$_\mathrm{outer}$ for the dusty sphere model and dusty shell models.
Given that the shape of the SED does not significantly change between Cycle 1 and Cycle 2, it is not unexpected that the temperatures of the dust components are roughly the same between epochs. 
If the dust is in the SN ejecta or pre-existing and actively interacting with the ejecta, we would expect some increase in the dust radius as the SN ejecta expands over the year between observations. Assuming an ejecta velocity of 7000 km/s (an upper estimate from the most recent spectrum, see Figure \ref{fig:o_lines}) in the 370 days between the Cycle 1 and Cycle 2 observations the ejecta radius should expand $\sim300 \times 10^3 R_\odot$ which, when accounting for errors, is consistent with the evolution between R$_\mathrm{outer}$ in Cycle 1 and Cycle 2 for the dusty sphere model and dusty shell models.

The difference in M$_\mathrm{hot}$ between epochs is likely due to the lack of constraints on the hot dust component in Cycle 1. Excluding F770W and the NIR photometry from the Cycle 2 fits produces M$_\mathrm{hot}$ values consistent with those observed in Cycle 1. 
This highlights the need for NIR photometry for constraining dust masses, particularly around SNe younger and hotter than SN~2017eaw. 
We note that there is no clear indication, in any of the three models, that the mass of the hot or cool dust components increased in the year between observations.  

The Cycle 1 JWST observations were previously modeled for the optically thin and dusty sphere case in \citet{Shahbandeh23}. \citet{Shahbandeh23} report only a total dust mass rather than separate mass components as we do here. However, we are able to reproduce all the dust properties for the dusty sphere of silicate dust case reported in \citet{Shahbandeh23} by assuming $M_\mathrm{tot} = M_\mathrm{hot}+M_\mathrm{cool} \approx M_\mathrm{cool}$ since $M_\mathrm{hot}<<M_\mathrm{cool}$. For the optically thin case, we are unable to replicate the temperature of the hotter dust component as \citet{Shahbandeh23}, though we reproduce the mass of the dust (again assuming $M_\mathrm{tot} \approx M_\mathrm{cool}$). However, we find that our Cycle 1 $T_\mathrm{hot}$ value is consistent within error with our Cycle 2 value, which is well constrained by the NIR data. {We note that there are no NIR spectra of SN~2017eaw at this epoch and therefore we cannot separate the SN's NIR line emission from the hot dust emission. Given this, the temperature of the hot dust component in our Cycle 2 models should be considered an upper limit.}

\subsection{Dust Geometry} \label{sec:geometry}
One of the primary ways to distinguish between newly-formed and pre-existing dust is to compare the geometry of the dust shell to that of the ejecta. The majority of pre-existing dust will not survive the interaction with the forward shock and ejecta and therefore cannot be located within the ejecta. Therefore, if the dust shell is at a radius sufficiently interior to the outermost ejecta, the dust must be newly-formed. Similarly, if the dust is outside the outermost radius of the ejecta then it must be pre-existing. 

First, to determine the robustness of any modeled dust radii measurements, we apply the same check as \citet{Zsiros24} and calculate the optical depth as follows \citep{Lucy89}:
\begin{equation}\label{eq:optdep}
    \tau = \kappa_\mathrm{average} \frac{M_\mathrm{dust}}{4\pi r^2},
\end{equation}
where $\kappa_\mathrm{average} = 750$ cm$^2$ g$^{-1}$ estimated for 0.1 $\mu$m silicate dust \citep[from grain properties in][]{Draine07, Sarangi22}. In the case of optically thin dust, the minimum outer radius of the dust is set by the blackbody radius. The dusty sphere case produces the minimum blackbody radius which is $R_{BB} = 3.39\times10^{5} R_\odot$ and $3.05\times10^{5} R_\odot$ for Cycle 1 and Cycle 2 respectively. If we assume this radius and a total dust mass of $5.4\times10^{-4}\ \mathrm{M}_\odot$ (based on the dust mass from the optically thin dust model), Equation \ref{eq:optdep} gives $\tau = 0.1$. Therefore, the observable dust around SN~2017eaw may be optically thin. 

We find the quality of the optically thin model fit to be comparable to that of the dusty sphere and dusty shell model fits for both epochs, further indicating that the dust may be optically thin. We also find that the dusty shell model converges in the case where the outer radius is set to be inside the ejecta radius  ($2\times10^{6}\ \mathrm{R}_\odot$, i.e. the radius at 2300 days assuming a velocity of 7000 km/s) and also does so when the inner radius is set to be greater than the ejecta radius. Both of these scenarios result in similar dust masses and temperatures for both dust components. We, therefore, assume that the dust is optically thin enough that the radius of the dust cannot be well constrained. Nevertheless, we compare the measured dust radii in the dusty sphere and shell models to the ejecta for completeness.

The radius at which dust resides in the ejecta is often quoted as a velocity coordinate with respect to the ejecta in order to account for the fact that the ejecta, and anything within it, is expanding over time. The outer edge of the ejecta is at a velocity of $\sim7000$\ km/s, given that the line profiles in the 1811 day and 2888 day nebular spectra indicate the outermost ejecta is interacting with CSM at this radius \citep[see Section \ref{sec:spectra} and ][]{Shahbandeh23}. To determine the velocity coordinate of the dust, we assume the simplest case $R = vt$, where $t$ is the time since explosion, $R$ is the radius of the dust, and $v$ is the velocity. If we use the radius derived from the dusty sphere model (since $R_\mathrm{sphere} < R_\mathrm{outer,shell}$), we find that the velocity coordinate of the dust emission is $2900^{+800}_{-400}$ km/s and $5000 \pm 1000$ km/s in Cycle 1 ($t\sim1960$ days) and Cycle 2 ($t\sim2330$ days), respectively. 

The radius within which dust formation can begin is the subject of debate.
Recent work by \citet{Sarangi22} suggests dust formation is confined to the 2500 km/s velocity coordinate, at least for the first 3000 days. Other studies have suggested this velocity coordinate may be as high as 5000 km/s \citep{Truelove1999, Maguire2012}. 
If we assume the radius from the dusty sphere model, the Cycle 1 dust component is near the region of the ejecta where dust formation may occur. The Cycle 2 dust component is consistent with dust both inside and outside the ejecta. For the dusty shell model, the velocity coordinate for Cycle 1 and 2 are $4500_{-800}^{+2500}$ and $9000^{+4000}_{-3000}$ km/s respectively, both of which are consistent with dust located outside of the ejecta. Regardless of model, the radii from the dust modeling could account for both pre-existing and newly-formed dust geometries.

\begin{figure*}[!]
    \centering
    \includegraphics[width=\hsize]{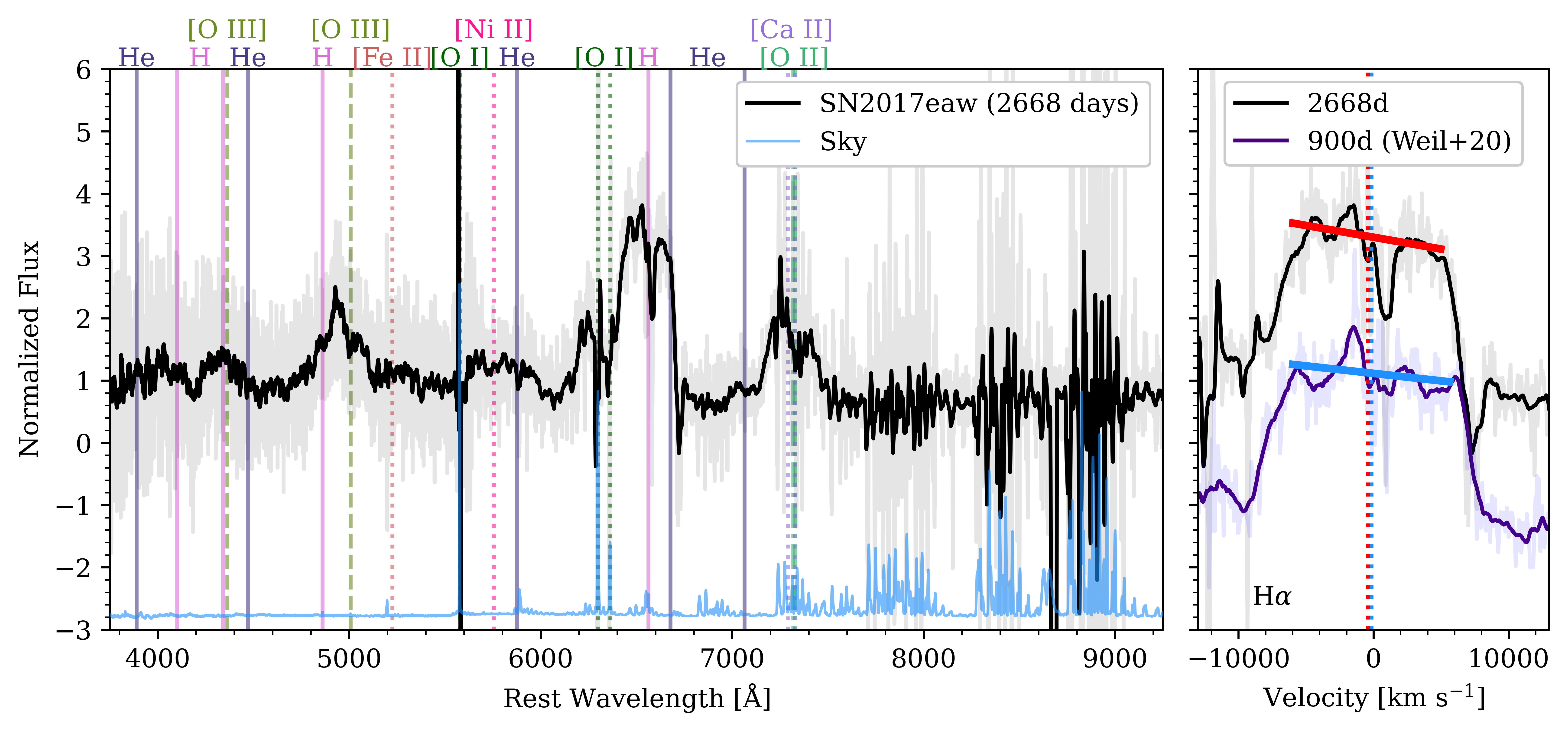}
    \caption{Left: The complete Keck LRIS spectrum taken at 2668 days post-explosion. The spectrum has been smoothed for clarity, the unsmoothed spectrum is displayed at lower opacity. The sky spectrum is shown below the SN~2017eaw spectrum for reference. Right: The H$\alpha$ profile of the most recent Keck spectrum compared to the 900 day spectrum published in \citet{Weil20}. The solid red and blue lines are a fit to the top of the 2668 and 900 day H$\alpha$ line profiles, respectively. The center of the profiles is marked by a line of the same color. 
    }
    \label{fig:spec}
\end{figure*}

\subsection{Spectral Evolution}\label{sec:spectra}

As shown in Figure \ref{fig:spec}, the nebular spectrum of SN~2017eaw continues to exhibit a prominent broad boxy H$\alpha$ profile, denoting continued CSM-ejecta interaction even at 2668 days after explosion. 
A boxy H$\alpha$ profile was first observed in SN~2017eaw 900 days post explosion \citep{Weil20}. A comparison of the 2668 day and 900 day H$\alpha$ profiles reveals that the center of the 2668 day profile is perhaps somewhat more blueshifted, $-430\pm40$ km/s compared to $-160\pm30$ km/s,  with a slightly steeper slope at the top of the line. This shape indicates dust attenuation, since the light from the receding ejecta, i.e. the red side of the line profile, is absorbed by the dust along the line of sight. However, it is difficult to robustly determine the impact of the dust attenuation given the signal to noise of the spectrum. The velocity of the ejecta, as measured at the location of full width half maximum, has decreased, 8000 km/s at 900 days compared to 7000 km/s at 2668 days, but some slowing is expected given the extent of the continued CSM interaction \citep{Dessart2024}. Ultimately, the H$\alpha$ at 2668 days is remarkably similar to that at 900 days and minimal evolution seems to have occurred in almost 2000 days. 

\begin{figure}
    \centering
    \includegraphics[width=\hsize]{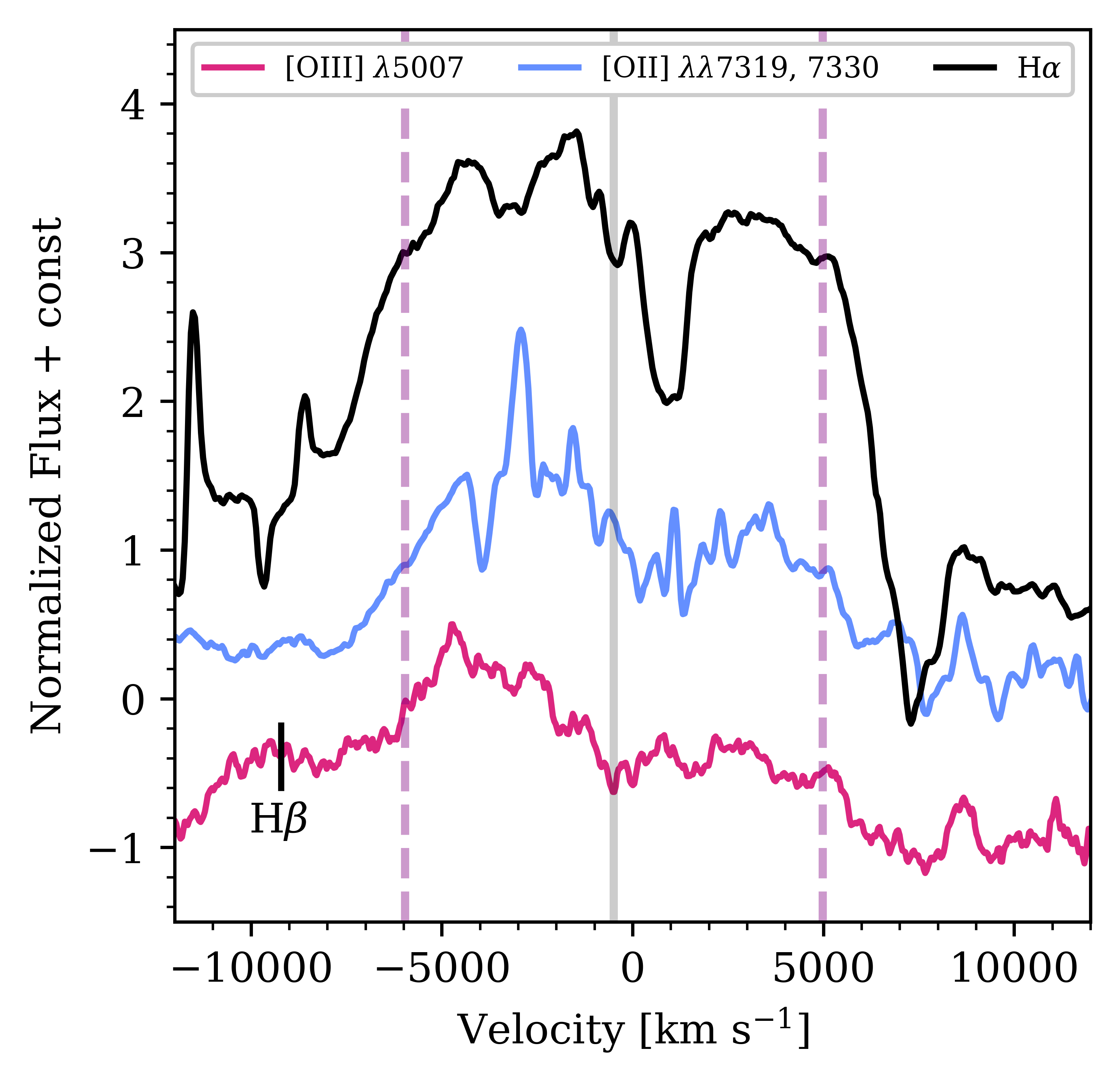}
    \caption{The line profiles of the three most prominent lines in the 2668 day Keck spectrum: [\ion{O}{3}] 
    $\lambda$5007, [\ion{O}{2}] $\lambda\lambda$7319,7330, and H$\alpha$. The purple dashed lines mark the edges of the top of the boxy H$\alpha$ profile. The grey line marks the center of the H$\alpha$ profile. The spectrum is smoothed for clarity.}
    \label{fig:o_lines}
\end{figure}

The three most prominent lines in the 2668 day Keck spectrum are H$\alpha$, [\ion{O}{2}] $\lambda\lambda7319,7330$, and [\ion{O}{3}] $\lambda5007$. We compare the profiles of these lines in Figure \ref{fig:o_lines}. We treat the [\ion{O}{2}] doublet as a line centered at 7324.5\AA. A [\ion{Ca}{2}] doublet can be present in the [\ion{O}{2}] line complex but the calcium doublet is clearly subdominant to the [\ion{O}{2}] lines at this epoch. 
Despite the existence of the [\ion{O}{3}] $\lambda\lambda4959, 5007$ doublet, the shape of the line profile in the 2668 day spectrum suggest the majority of the light is from the stronger of the lines, we therefore attribute the entire profile to [\ion{O}{3}] $\lambda5007$. The [\ion{O}{3}] $\lambda5007$ line complex also contains H$\beta$, which is visible on the blue shoulder.  Again, [\ion{O}{3}] $\lambda5007$ is the stronger of the two lines. Therefore we treat the [\ion{O}{2}] and [\ion{O}{3}] lines as primarily oxygen in our analysis. 

The edges of the oxygen profiles line up remarkably well with the edges of the H$\alpha$ profile. The oxygen lines are notably attenuated on the red-side of the profile. This effect is less pronounced in the hydrogen line but still seems to be present. This might suggest there is a reservoir of dust inside the ejecta which is absorbing light from the far side of the supernova. If this attenuation is due to newly formed dust in the ejecta, rather than some asymmetry in the explosion and/or CSM, the strength of the attenuation and the low dust mass revealed in the JWST/MIRI images suggests that the majority of this interior dust is likely too optically thick to observe in the mid-infrared. 

Recent work by \citet{Dessart2025} presents models of a {SN IIP} at 1000 days post-explosion that is CSM-interacting and also contains a small mass of dust in the cold dense shell. They find there is no significant dust attenuation effect for dust interior to the ejecta due to the small angle of the inner dust relative to the emitting region of the outer ejecta. Further, in contrast to non-interacting SNe, in the \citet{Dessart2025} CSM-interacting SN model interior dust has no impact on the line strengths, or the hydrogen-oxygen line ratios, due to the fact that 99.7\% of the model emission is from the outer ejecta. Indeed, \citet{Weil20} note no significant signs of dust attenuation in SN~2017eaw at 900 days post-explosion. At 2668 days post-explosion, the velocity of the ejecta (7000 km/s) is similar to that of the 1000 day model (8000 km/s) in \citet{Dessart2025} and it is possible that the observed line attenuation is not due to interior dust. Although this physical picture may still be valid, we caution against direct comparisons given that both the inner and outer regions have evolved for an additional $\sim$1700 days. The velocity of the ejecta at 2668 days suggests that the outer regions of the ejecta have somewhat slowed due to CSM interaction and the conclusions from $~$1000 days may no longer be valid. \citet{Dessart2025} also investigate the effect of $10^{-4}$, $5\times10^{-4}$, and $10^{-3}\ \mathrm{M_\odot}$ of dust in the ejecta at a velocity coordinate of 8000 km/s. 
These models produce H$\alpha$ profiles very similar to the one at 2668 days. The model profiles include a dip in the center (near rest wavelengths) due to the increased optical depth of the limbs of the shell. We refrain from definitively linking a similar feature in the observed H$\alpha$ profile with dust given the low SNR of the spectrum.
Further modeling of the effects of dust in CSM-interacting SNe II at late times ($>$1000 days post-explosion) is needed to understand the evolution of H$\alpha$ and the strongest oxygen lines. 

\section{Dust Origin Scenarios} \label{sec:discussion}
 
Despite the general decrease in mid-infrared flux from Cycle 1 to Cycle 2, the SED evolution of SN~2017eaw points to dust which has not significantly changed in the year between JWST/MIRI observations. Given the optically thin nature of the dust and the constant dust temperatures, this drop in flux could be due to a decline in dust mass. These observations indicate a possible trend to watch for in future observations and set a robust baseline against which future measurements can be compared. Even if the dust mass is not decreasing, it has not increased as might be expected if dust formation was actively occurring. Further, the lack of temperature evolution indicates that the mass of any newly-formed dust that cooled between the JWST observations is small compared to the total dust mass observable in the mid-infrared. 

As was noted in \citet{Shahbandeh23}, even the cooler component of dust is too warm to be heated only by the ejecta of SN~2017eaw. The observed dust temperatures require an external heating mechanism to be present. Given that neither of the components have cooled or heated markedly since the initial observations reported in \citet{Shahbandeh23}, the external heating mechanism must be maintained over the course of the year between observations. The possible heating mechanisms depend on the location of the dust and can therefore be used to probe the dust origin. We explore three different scenarios where the mid-infrared dust is primarily 1) pre-existing in the CSM and collisionally heated; 2) pre-existing and radiatively heated, and 3) newly-formed in the ejecta and radiatively heated. 

\subsection{Collisionally Heated Pre-existing Dust}

\begin{figure}
    \centering
    \includegraphics[width=\hsize]{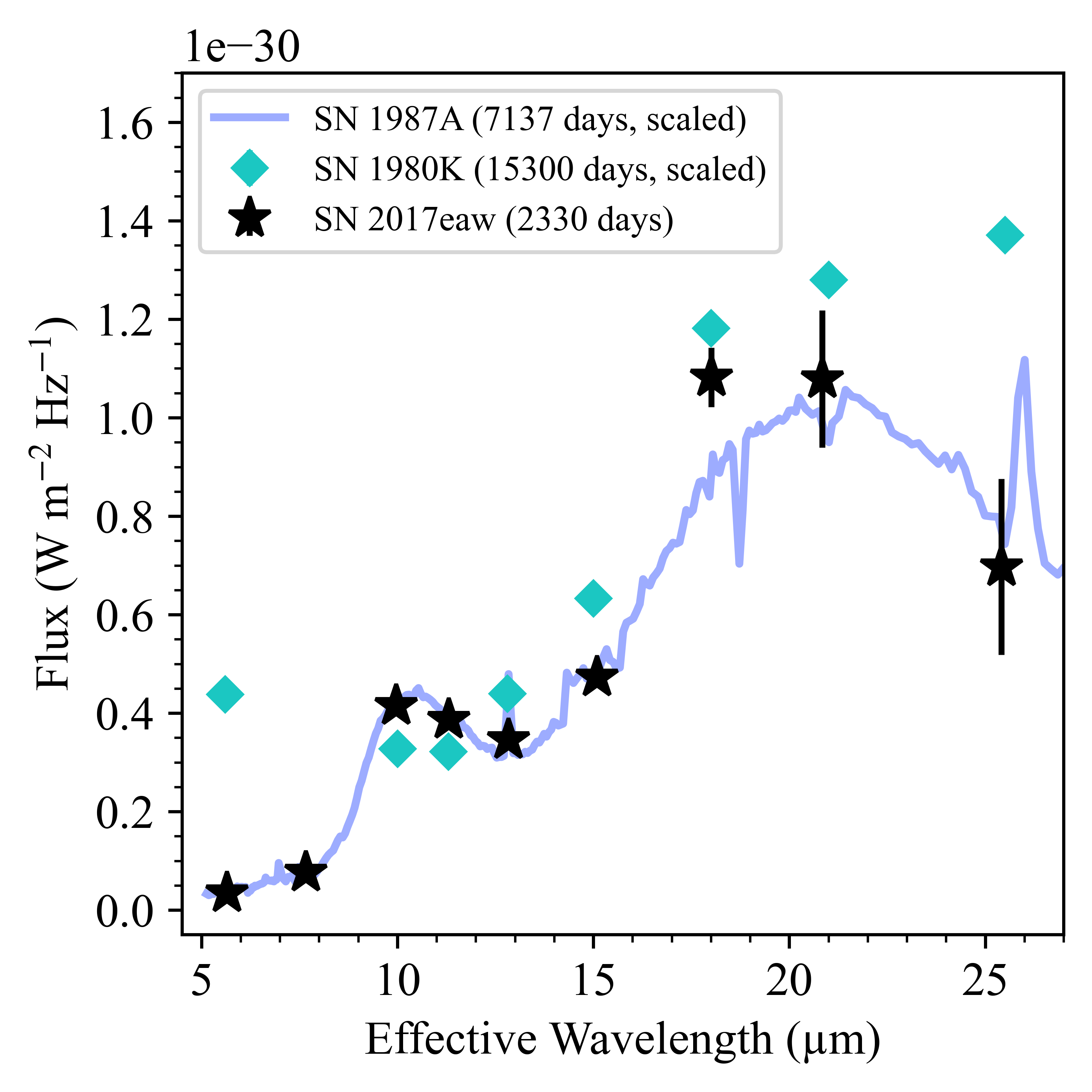}
    \caption{A comparison of SN~2017eaw's mid-infrared SED to the mid-infrared SED of SN~1980K \citep{Zsiros24} and a Spitzer IRS spectrum of SN~1987A \citep{Dwek10}. The SN~1987A and SN~1980K data have been scaled to the flux of the SN~2017eaw SED from 10-15 \micron. The SN~2017eaw SED is strikingly similar particularly to that of SN~1987A at 6000-8000 days. The mid-infrared flux during this epoch of SN~1987A's evolution is likely due to cool pre-existing dust collisionally heated by the strong interaction between the SN ejecta and the equatorial ring \citep{Arendt16}.}
    \label{fig:87A_80K_comp}
\end{figure}

The mid-infrared SED of SN~2017eaw is somewhat similar in shape to that of SNe 1987A and 1980K at significantly later phases, see Figure \ref{fig:87A_80K_comp}. In the case of both SNe 1987A and 1980K, the 8-20 $\mu$m emission is consistent with 150-180K silicate dust, remarkably similar to SN~2017eaw. Notably the mid-infrared flux in SN~2017eaw, which is primarily dominated by the cool dust component, is almost identical in shape to Spitzer IRS spectra of SN~1987A between 6000 and 8000 days post-explosion. This component in SN~1987A has been linked to the collisional heating of the equatorial ring \citep{Dwek10, Arendt16, Arendt20}, and a similar scenario was suggested for SN~1980K \citep{Zsiros24}. The mass of dust in the $\sim$160K component of SN~2017eaw is $\sim$1 order of magnitude larger than observed in SN~1987A and $\sim$1 order of magnitude smaller than observed in SN~1980K. 

To determine if it is plausible for the cooler component of SN~2017eaw's mid-infrared dust to be collisionally heated via interaction between the ejecta and CSM, we follow the method of \citet{Fox10} for estimating the mass of dust processed by the forward shock \citep[see also][]{Fox11, Tinyanont16, Zsiros22, Zsiros24}. Any pre-existing dust must reside outside the evaporation radius, inside which the peak luminosity of the SN will have destroyed any pre-existing dust grains. Assuming the temperature and peak bolometric luminosity measured by \citet{Szalai19}, 14,000K and $\sim$10$^{43}$ erg/s (rounded from L$_\mathrm{peak}\approx 7\times 10^{42}$ erg/s) respectively, the evaporation radius (R$_\mathrm{evap}$) is $2.7\times10^4$ R$_\odot$. R$_\mathrm{evap}$ is significantly less than the ejecta radius at 2,000 days. In both JWST epochs, the SN ejecta has far surpassed the evaporation radius and could feasibly be interacting with CSM containing pre-existing dust. 

In the case of collisional heating, the hot, post-shocked gas heats a shell of pre-existing dust. The total mass of this dust can be determined from the volume of the emitting shell using equations for grain sputtering and by assuming a dust-to-gas ratio of 0.01 \citep{Fox10}. This gives
\begin{equation}\label{eq:collision}
    \frac{M_{\mathrm{dust}}}{M_{\odot}} \approx 0.0028 \left(\frac{\nu_\mathrm{s}}{15,000\ \mathrm{km\ s}^{-1}}\right)^3 \left(\frac{t}{\mathrm{year}}\right)^2 \left(\frac{a}{\mu \mathrm{m}}\right),
\end{equation}
where $\nu_\mathrm{s}$ is the shock velocity, $t$ is the time since explosion, and $a$ is the dust grain size. Similar to \citet{Zsiros24}, we use $\nu_\mathrm{s}=$ 5,000 km s$^{-1}$ and 15,000 km s$^{-1}$ and $a=$ 0.005 and 0.1 $\mu$m (we assume $a=0.1$ $\mu$m in our dust modeling) as our lower and upper bounds, respectively. Assuming these values, the range of dust masses that could be collisionally heated is $\sim10^{-5} - 10^{-2} M_\odot$. The total dust masses observed in both Cycle 1 and 2 are in the middle of this range. We note that the velocity of the ejecta measured from the nebular spectra, 7000 km s$^{-1}$, with a 0.1 $\mu$m dust grain could result in a collisionally heated dust mass of $10^{-3}\ M_\odot$. This result suggests that the majority of the dust observed in the mid-infrared could reasonably be collisionally heated pre-existing dust. 

\begin{figure}
    \centering
    \includegraphics[width=\hsize]{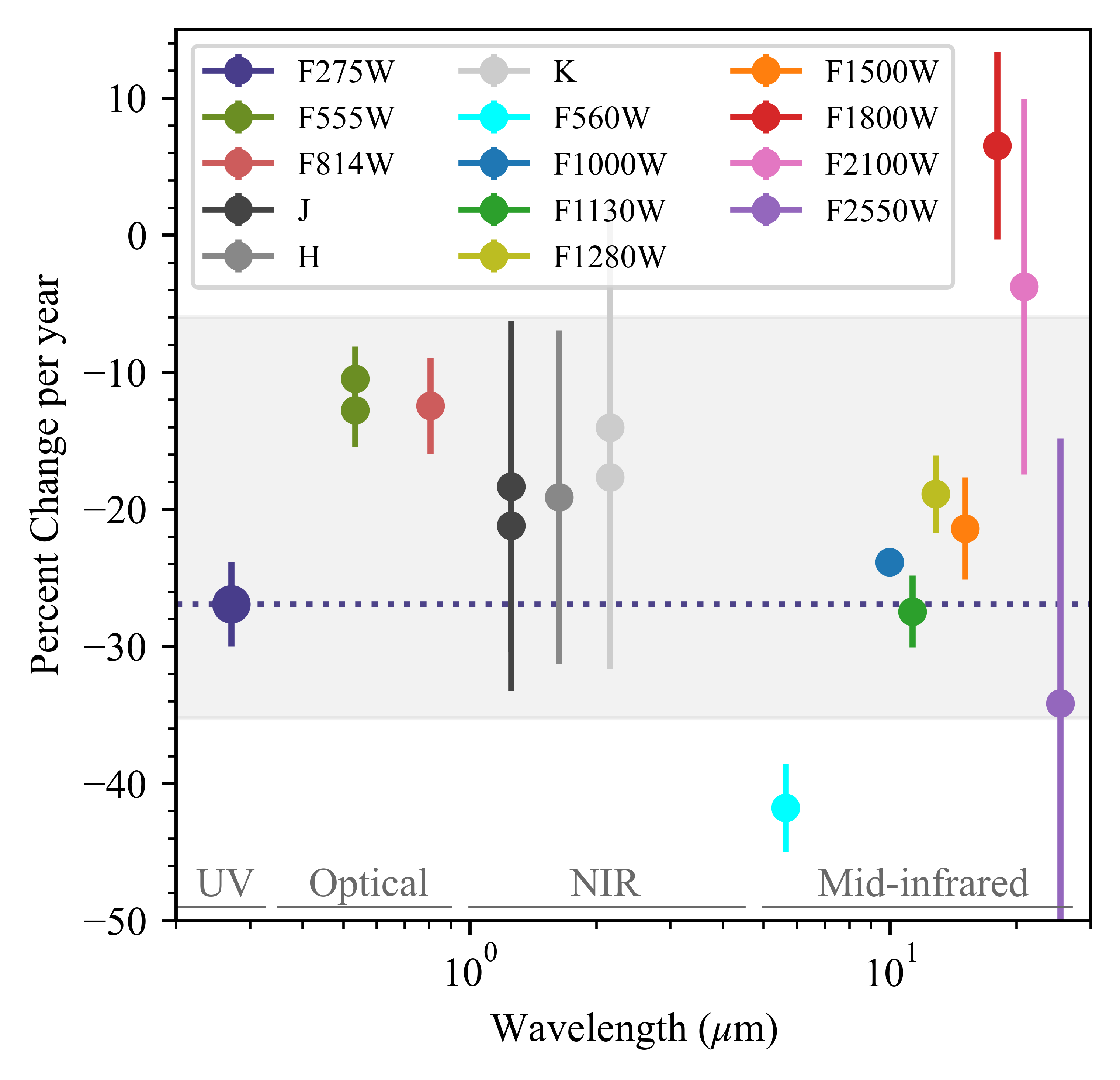}
    \caption{The percent change in flux per year for SN~2017eaw. The UV flux of SN~2017eaw decreased from 2020 to 2023, indicating that the observed UV excess is likely the result of CSM interaction rather than an underlying star or stellar population. The grey region denotes 1 standard deviation from average mid-infrared flux change, this lines up well with the observed decrease in both the NIR and UV (also plotted as a navy dotted line for reference). This suggests the mid-infrared, NIR, and UV are probing a similar region of the SN ejecta.}
    \label{fig:fluxchange}
\end{figure}

The correlation between the UV and mid-infrared flux evolution might also suggest the observed dust is collisionally heated. 
As shown in Figure \ref{fig:fluxchange}, the flux in F275W decreased 27\% per year between 2020 and 2023. Similarly, the MIRI flux decreased an average of 21\% ($\pm 15$\%) per year across all filters between the Cycle 1 (2022) and Cycle 2 (2023) observations. The F275W filter notably includes \ion{Mg}{2} $\lambda\lambda$ 2796, 2803, one of the UV features most strongly affected by CSM interaction \citep{Dessart23}. Therefore the evolution in F275W can be used as a proxy for the extent of CSM. The drop in F275W flux and corresponding drop in the mid-infrared suggests that both wavelength regimes are probing the same medium. This correlation between UV and mid-infrared flux is similar to the X-ray (which similarly probes CSM interaction) and mid-infrared evolution observed in SN~1987A at 6000-8000 days, during which the IR-to-X-ray flux ratio remains constant. As shown in Figure \ref{fig:87A_80K_comp}, the mid-infrared SED of SN~2017eaw is nearly identical to that of SN~1987A during this epoch where the 8-20 $\mu$m dust component is believed to be collisionally heated dust in the equatorial ring \citep{Dwek10, Arendt16}. 

Furthermore, the reduction in UV flux does not correlate with a change in the temperature of the dust. If the dust, whether pre-existing or newly formed, is radiatively heated by CSM interaction, the temperature of the dust is expected to decrease with the UV flux and therefore CSM interaction. There is no evidence of the majority of the mid-infrared dust cooling between epochs.

\subsection{Radiatively Heated Pre-existing Dust}
The theory that the cool dust component is collisionally heated assumes a linear decline in UV luminosity from 2020 to 2023, but there is no UV data between these epochs to track the decline. It is possible that the UV luminosity was constant enough in the year between the JWST/MIRI observations for the dust temperature to not substantially change over the course of the year. In this scenario, radiative heating could still account for the lack of temperature evolution in the dust. The decrease in mid-infrared flux could be attributed to a changing geometry of the dust shell illuminated by the CSM interaction. We therefore can not use the similar UV and mid-infrared decline rates to completely rule out the possibility of radiative heating.

A simple {infrared} echo model, assuming the light from the SN explosion excites pre-existing dust, places the echo radius at $R_\mathrm{echo} = ct_\mathrm{echo}/2$, where $t_\mathrm{echo}$ is the duration of the light echo \citep{Bode80, Dwek83}. Assuming a lower limit of $t_\mathrm{echo} = 2330$ days in order for the echo to still be detectable in the Cycle 2 mid-infrared observations, this gives $R_\mathrm{echo} = 4.3\times10^7\ R_\odot$, significantly above the outer dust radii given by the non-optically thin models in Section \ref{sec:dustmod}. When the ejecta is between the evaporation and the light echo radii, as is the case for SN~2017eaw, the luminosity from CSM-ejecta interaction heats the pre-existing grains creating a CSM echo \citep{Fox10, Fox11}\footnote{Sometimes referred to as a circumstellar shock echo}. \citet{Shahbandeh23} found that for SN~2017eaw the optical luminosity necessary to heat the grains to the temperature of the cool dust component in the Cycle 1 data exceeds the observed optical luminosity. Unsurprisingly, we find this to be the case for the Cycle 2 dust as well. 

However, \citet{Dessart22} suggests that CSM-ejecta interaction may primarily produce UV emission, especially in Ly$\alpha$ and Mg II $\lambda\lambda$ 2796, 2803. 
Assuming constant luminosity across the UV ($10-400$ nm), we use the F275W observation to estimate $\mathrm{L}_\mathrm{UV} = 3\times10^5\ \mathrm{L}_\odot$ in 2023. This may be an overestimation of the total UV luminosity given the Mg II lines fall into the F275W filter. Nevertheless, this value is similar to the $\sim10^5 \mathrm{L}_\odot$ modeled in \citet{Dessart22}, indicating that the UV luminosity could account for the temperature of the observed mid-infrared dust. More extensive wavelength coverage in the UV is required to place robust limits on the UV luminosity.

\subsection{Newly-formed Dust}
Pre-existing dust does not preclude the existence of newly-formed dust in the ejecta. The detection of CO in the ejecta between $200-500$ days post explosion indicates that the ejecta has long been cool enough to form dust \citep{Tinyanont19}. 
Further, there are signs of blue-shifted line profiles in the spectra at roughly 2000 days (just before the JWST Cycle 1 observations) \citep{Shahbandeh23} and at 2668 days (as discussed in Section \ref{sec:spectra}). The observed dust attenuation in the red side of the line profiles suggests there is dust in the ejecta of the SN. 

The optical spectra of SN~2017eaw indicate that the SN is producing dust \citep[although see][]{Dessart2025}. However, the newly-formed dust might be too optically thick to be observed in the mid-infrared even at $>2000$ days post explosion. In this case, only thermal emission from the outermost shell of the total mass of newly-formed dust will be observable. This outermost layer would only constitute an extremely small percentage of the total newly-formed dust mass and may not noticeably change the mid-infrared SED if a more massive amount of pre-existing dust is also present. 
Nevertheless, geometrical arguments (see Section \ref{sec:geometry}) indicate that some of the observed mid-infrared dust could be located within the ejecta or in a cold dense shell between the forward and reverse shock. Given the optical depth and evolution of the cool dust component, it is unlikely to be primarily newly-formed though some component of this dust may be. 

In contrast, there are minimal constraints on the evolution of the hot dust component due to the lack of NIR observations during Cycle 1, and there may have been dust growth and/or cooling between the Cycle 1 and 2 observations as would be expected of a newly formed dust component. There is no NIR spectra of SN~2017eaw at this late epoch so we are unable to quantify the extent to which the NIR emission is from the SN itself rather than dust. NIR spectra of SN~1987A at around 2000 days shows strong emission lines in $J$ band but none in $K$ band \citep{Fassia2002}. If this is also the case for SN~2017eaw, the hot dust component could be slightly cooler than modeled. Further, there is likely some flux in this component that is due to blackbody emission from the SN. Our inferred cool dust mass of $5.4\times10^{-8}$ $\mathrm{M}_\odot$ (see Table \ref{tab:dustparameters}) should be treated as a upper limit of the amount of dust found in the hot component. Given the small amount dust in the hot component (relative to the cold component), this uncertainty does not impact the conclusions of this work. 
Ultimately, we are unable to confirm the origin of the hot dust component from the existing observations. Further observations of SN~2017eaw should also include NIR observations to place constraints on the evolution of the hot component and provide insights into its origin.

\subsection{Implications for Dust Formation in CCSNe}\label{sec:dustevo}

\begin{figure*}
    \centering
    \includegraphics[width=\hsize]{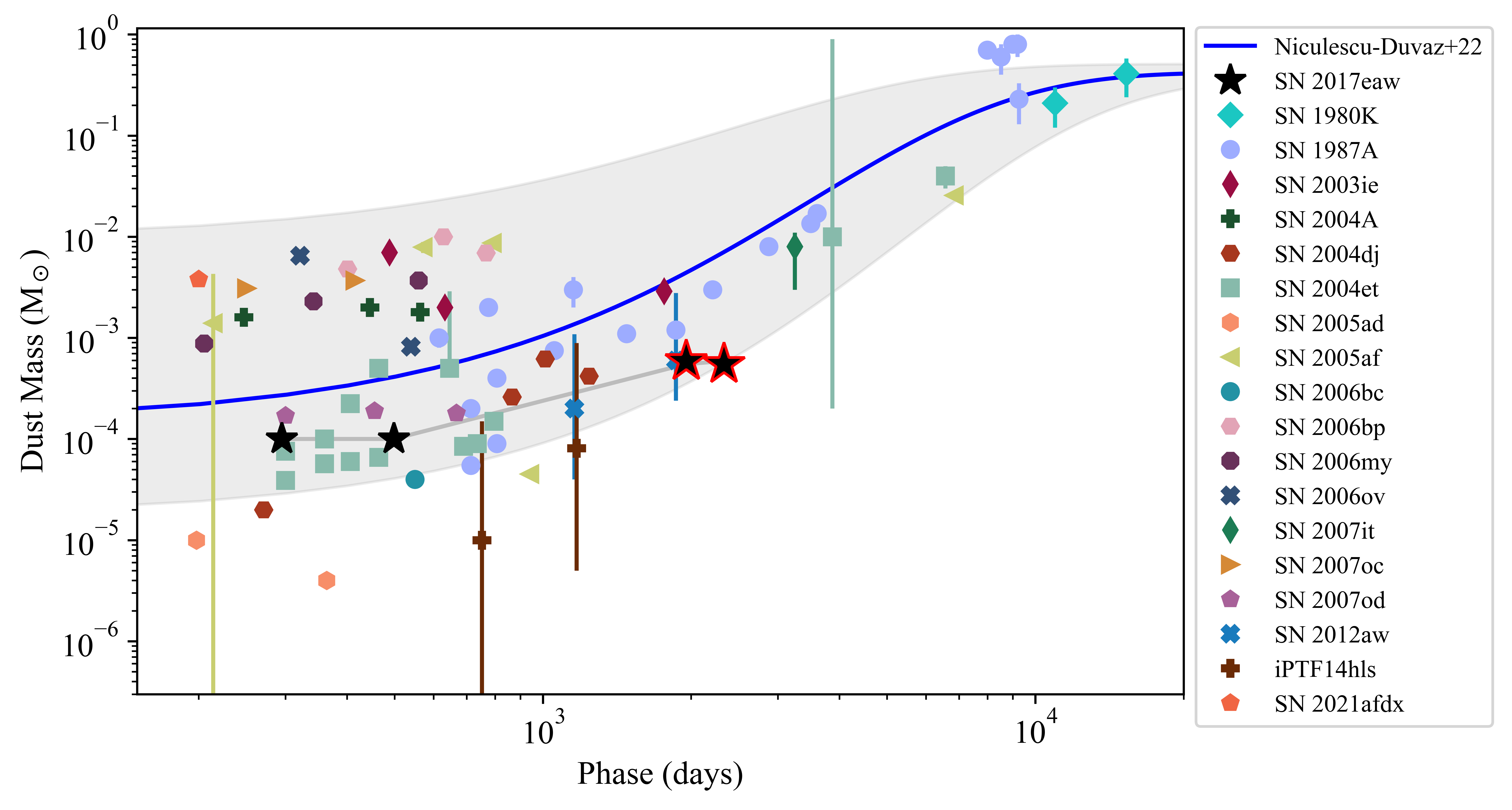}
    \caption{Dust mass versus time for a collection of Type II CCSNe. Dust masses for SN~2017eaw derived in this work (red-edged black stars) do not significantly differ from the dust production trend presented in \citet[][their Figure 23]{Niculescu-Duvaz2022} denoted by the blue line and grey shaded region. Supernova dust measurements in this figure include SN~2017eaw {\citep{Tinyanont19}}, SN~1980K {\citep{2017MNRAS.465.4044B,Zsiros24}}, SN~1987A {\citep{Matsuura11,Indebetouw14,Matsuura15,2015MNRAS.446.2089W,2016MNRAS.456.1269B}}, SN~2003ie {\citep{Szalai13}}, SN~2004A {\citep{Szalai13}}, SN~2004dj {\citep{2011A&A...527A..61S}}, SN~2004et {\citep{Kotak09,2011MNRAS.418.1285F,Niculescu-Duvaz2022,Shahbandeh23}}, SN~2005ad {\citep{Szalai13}}, SN~2005af {\citep{Szalai13, Sarangi2025}}, SN~2006bc {\citep{2012ApJ...753..109G}}, SN~2006bp {\citep{Szalai13}}, SN~2006my {\citep{Szalai13}}, SN~2006ov {\citep{Szalai13}}, SN~2007it {\citep{Niculescu-Duvaz2022}}, SN~2007oc {\citep{Szalai13}}, SN~2007od {\citep{2010ApJ...715..541A}}, SN~2012aw {\citep{2023MNRAS.519.2940N}}, iPTF14hls {\citep{2023MNRAS.519.2940N}}, and SN~2021afdx {\citep{Hosseinzadeh23}}.}
    \label{fig:dusttimeline}
\end{figure*}

Significant work has been done to understand the timeline of dust formation in CCSNe. When SN~2017eaw's mid-infrared dust mass is compared to literature values of CCSNe dust masses, it lies near the lower limit of the dust trend observed in \citet{Niculescu-Duvaz2022}, as shown in Figure \ref{fig:dusttimeline}. The slight fluctuation in dust mass from $\sim2000$ to $\sim2300$ days post-explosion is similar to trends observed in several other SNe, though this behavior has never been observed in another SN $>2000$ days post-explosion. However, SN~2017eaw is the only SN other than SN~1987A with multiple epochs of mid-infrared observations between 1000-5000 days post-explosion.

If the mid-infrared dust emission in SN~2017eaw is primarily due to pre-existing dust, then its location on the dust formation timeline may be significantly different than shown in Figure \ref{fig:dusttimeline}. It is possible that many of the early time mid-infrared dust measurements of CCSNe are similarly contaminated with pre-existing dust. Late time dust mass measurements of SN~1987A were done in the far-IR and sub-mm and probed dust significantly colder than can be observed with JWST \citep{Matsuura11, Indebetouw14, Matsuura15}. For SN~1980K, reported dust measurements were measured by modeling the dust attenuation on optical spectral lines. The mid-infrared SED of SN~1980K reveals 100 times less dust than indicated by the line profiles \citep{Zsiros24}. This suggests that the majority of the dust in SN~1980K is also too cold to be observed by JWST/MIRI. The same might be true for SN~2017eaw but the signal to noise of the recent spectra makes modeling of the line profiles difficult. 

However, SN~2017eaw is significantly younger than both SN~1980K and SN~1987A. Any newly-formed dust around SN~2017eaw should be more optically thick than observed in the two older supernovae. In the case of optically thick dust, radiative processes from CSM-ejecta interaction will only heat the outermost shell of newly-formed dust, which is likely to make up only a tiny amount of the total dust mass. Over the course of a year, the expansion of a shell of newly-formed dust may not be enough to visibly evolve the mid-infrared SED. 

\section{Summary \& Conclusion}\label{sec:summary}
We present late time UV, optical, and near-infrared observations of SN~2017eaw to map its spectral energy evolution. The SN has declined in flux across almost all wavelengths.
We find that the NIR flux has declined below the progenitor level, confirming the progenitor detection. SN~2017eaw is still detected in HST WFC3/UVIS F275W, and the optical spectrum at 2668 days exhibits broad boxy line profiles, particularly H$\alpha$, indicating that there is continued CSM-ejecta interaction even at $>$2500 days post-explosion. 

SN~2017eaw is one of the first supernovae to have multi-epoch JWST MIRI imaging. These observations reveal that the mid-infrared flux has decreased in most filters in the year between the MIRI observations. 
SED modeling reveals a hot ($\sim$1700~K) silicate dust component of $5.4\times10^{-8}\ \mathrm{M}_\odot$ and a cool ($\sim$160~K) silicate dust component of $5.5\times10^{-4}\ \mathrm{M}_\odot$ consistent with being optically thin. Here we cite the dust modeling values for the dusty shell case since these values are between the optically thin and dusty sphere models.
Interestingly, there is no indication that the dust is cooling or increasing in mass as might be expected for dust which is actively forming.

Furthermore, the decline in mid-infrared flux is similar to that observed in the UV, perhaps hinting that the dust observed in the mid-infrared is located in the same CSM whose interaction with the ejecta is producing UV flux. 
The evolution in the UV suggests a changing CSM density or geometry around SN~2017eaw. To understand how this continues to affect the dust budget and the dust heating mechanism, continued X-ray and UV observations are necessary. 

The multi-wavelength evolution of SN~2017eaw suggest that, while there may be newly-formed dust in the ejecta or cold dense shell, a significant fraction of the cool dust observed in the mid-infrared is likely pre-existing. 
There is a need for further late time ($>$1000 days post-explosion) multi-wavelength observations for the nearest supernovae, like SN~2017eaw, in order to map the extent and duration of CSM-interaction and its impact on dust evolution. Such observations, spanning the UV to the mid-infrared, will reveal insights into red supergiant mass loss in the final years before death and help to constrain the timeline of new dust production in SNe II. 

\section{Acknowledgments}
We thank R.G. Arendt for providing the Spitzer IRS spectra of SN~1987A. Thank you to C. DeCoursey and J. Pierel for their help with JWST photometric reduction using space\textunderscore phot. We also thank D. Perley for helpful advice in reducing the SN2017eaw Keck LRIS spectrum.

{The complete set of HST and JWST data presented in this article were obtained from the Mikulski Archive for Space Telescopes (MAST) at the Space Telescope Science Institute. The specific observations analyzed can be accessed via \dataset[doi: 10.17909/tk94-1430]{https://doi.org/10.17909/tk94-1430}.}

This work is based in part on observations made with the NASA/ESA/CSA James Webb Space Telescope. The data were obtained from {MAST} at the Space Telescope Science Institute, which is operated by the Association of Universities for Research in Astronomy, Inc., under NASA contract NAS 5-03127 for JWST. These observations are associated with program GO3295 and GO2666. 

This research is based in part on observations made with the NASA/ESA Hubble Space Telescope obtained from the Space Telescope Science Institute, which is operated by the Association of Universities for Research in Astronomy, Inc., under NASA contract NAS 5-26555. These observations are associated with programs SNAP17070 and SNAP17506. 

Observations reported here were obtained at the MMT Observatory, a joint facility of the University of Arizona and the Smithsonian Institution.

Some of the data presented herein were obtained at Keck Observatory, which is a private 501(c)3 non-profit organization operated as a scientific partnership among the California Institute of Technology, the University of California, and the National Aeronautics and Space Administration. The Observatory was made possible by the generous financial support of the W. M. Keck Foundation. 

The authors wish to recognize and acknowledge the very significant cultural role and reverence that the summit of Maunakea has always had within the Native Hawaiian community. We are most fortunate to have the opportunity to conduct observations from this mountain.

JEA is supported by the international Gemini Observatory, a program of NSF NOIRLab, which is managed by the Association of Universities for Research in Astronomy (AURA) under a cooperative agreement with the U.S. National Science Foundation, on behalf of the Gemini partnership of Argentina, Brazil, Canada, Chile, the Republic of Korea, and the United States of America.

Time domain research by the University of Arizona team and D.J.S. is supported by National Science Foundation (NSF) grants 2108032, 2308181, 2407566, and 2432036 and the Heising-Simons Foundation under grant $\#$2020-1864.

CDK gratefully acknowledges support from the NSF through AST-2432037, the HST Guest Observer Program through HST-SNAP-17070 and HST-GO-17706, and from JWST Archival Research through JWST-AR-6241 and JWST-AR-5441.

W.J.-G. is supported by NASA through Hubble Fellowship grant HSTHF2-51558.001-A awarded by the Space Telescope Science Institute, which is operated for NASA by the Association of Universities for Research in Astronomy, Inc., under contract NAS5-26555.

AAM, CL, and NR are supported by DoE award no.~DE-SC0025599. MMT Observatory access for AM, CL, and NR was supported by Northwestern University and the Center for Interdisciplinary Exploration and Research in Astrophysics (CIERA).

KAB is supported by an LSST-DA Catalyst Fellowship; this publication was thus made possible through the support of Grant 62192 from the John Templeton Foundation to LSST-DA.

\facilities{HST (ACS, WFC3), JWST (MIRI), Keck (LRIS), MAST (HLSP), MMT (Binospec, MMIRS)}

\software{\texttt{astropy} \citep{AstropyCollaboration2013, AstropyCollaboration2018, AstropyCollaboration2022}, Dolphot \citep{Dolphin00, Dolphin16}, emcee \citep{emcee}, Light Curve Fitting \citep{lightcurvefitting, lightcurvefitting2}, LPipe \citep{Perley2019}, MatPLOTLIB \citep{mpl}, NumPy \citep{numpy}, Photutils \citep{photutils}, \texttt{space\textunderscore phot} \citep{spacephot}, SEP \citep{Sextractor, SEP}, WebbPSF \citep{WebbPSF1, WebbPSF2}, WISeREP \citep{wiserep}}

\bibliography{sources}
\bibliographystyle{aasjournalv7}

\appendix
\label{appendix}

\begin{figure}
    \centering
    \includegraphics[width=\hsize]{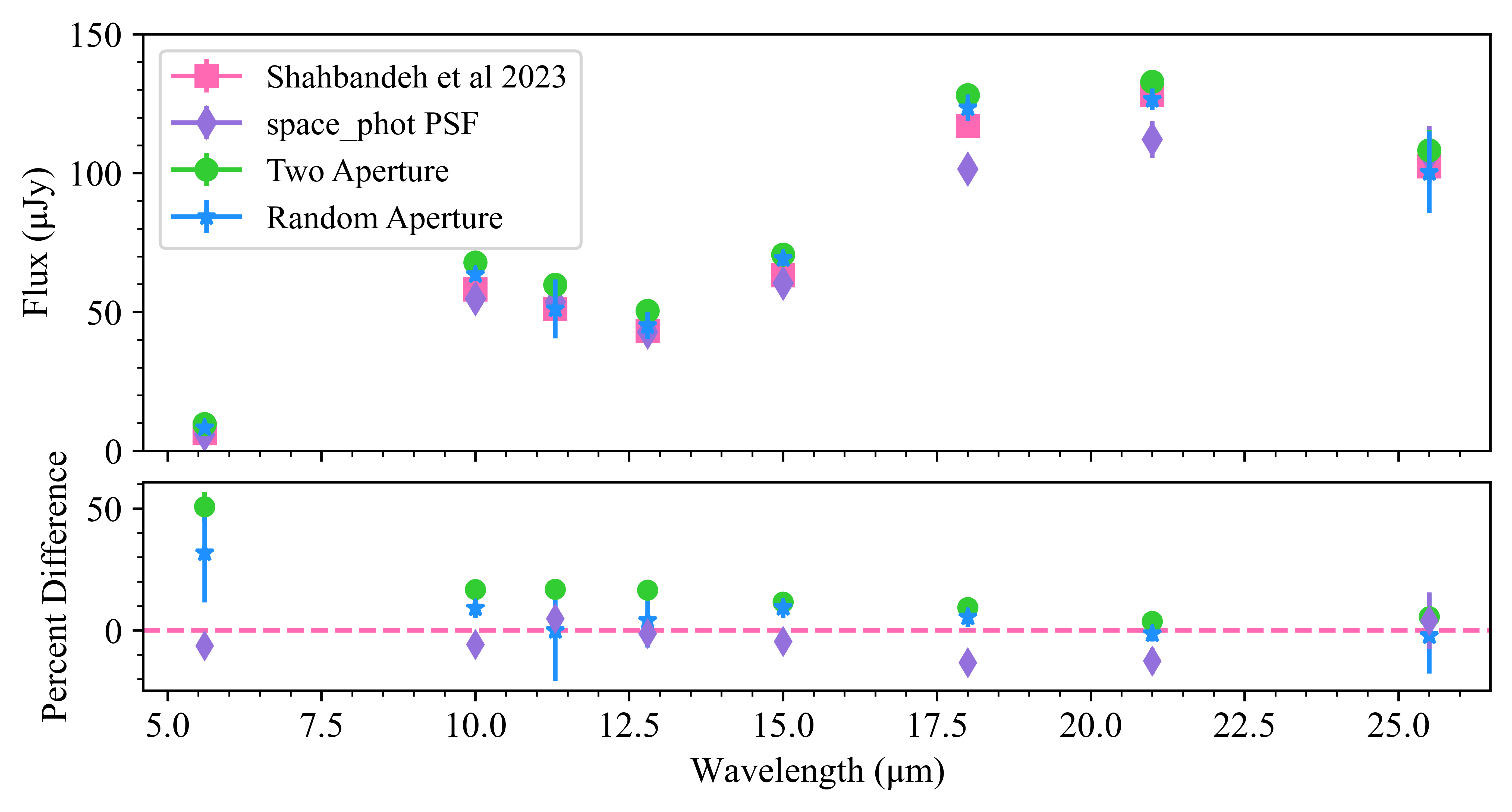}
    \caption{Comparison of aperture and PSF photometry methods for JWST/MIRI Cycle 1 observations of SN~2017eaw. Top: JWST/MIRI SED measured using the different photometric methods. Also plotted are the published values from \citet{Shahbandeh23}. The aperture methods result in fluxes which are higher than those reported for PSF methods. Bottom: Percent difference in flux from \citet{Shahbandeh23} values for each photometry method in this work. The \texttt{space\textunderscore phot} method is consistent with the \citet{Shahbandeh23} values, therefore we report this photometry in Section \ref{sec:jwstphot}. We note that significant changes were made to the MIRI PSFs (primarily $>$15 $\mu$m) following in the publication of \citet{Shahbandeh23}, this is likely the cause of the discrepancy between the \texttt{space\textunderscore phot} and the published values for the redder filters.}
    \label{fig:ap_vs_psf}
\end{figure}

\section{Apperture vs. PSF Photometry}\label{sec:ap_v_psf}

In order to measure the flux in the MIRI images, we attempted several different photometric techniques which we compared to the published Cycle 1 photometry in \citet{Shahbandeh23}. In this work, we report JWST MIRI PSF photometry of SN~2017eaw as this methodology resulted in values that are most consistent with previously published photometry. \citet{Shahbandeh23} did PSF photometry on the stage 2 products, a method similar to the one we use for our photometry in Section \ref{sec:jwstphot}. As shown in Figure \ref{fig:ap_vs_psf} (purple diamonds and pink squares), the most significant offsets between our \texttt{space\textunderscore phot} PSF photometry and the published photometry are at 18 and 21 \micron, both of which are filters where the photometric calibration maps were significantly updated\footnote{This work uses version 0056, details can be found at \url{https://jwst-crds.stsci.edu/browse/jwst_miri_photom_0056.rmap}} between the publication of \citet{Shahbandeh23} and the completion of this work, therefore this offset is unsurprising.

\begin{figure}
    \centering
    \includegraphics[width=\hsize]{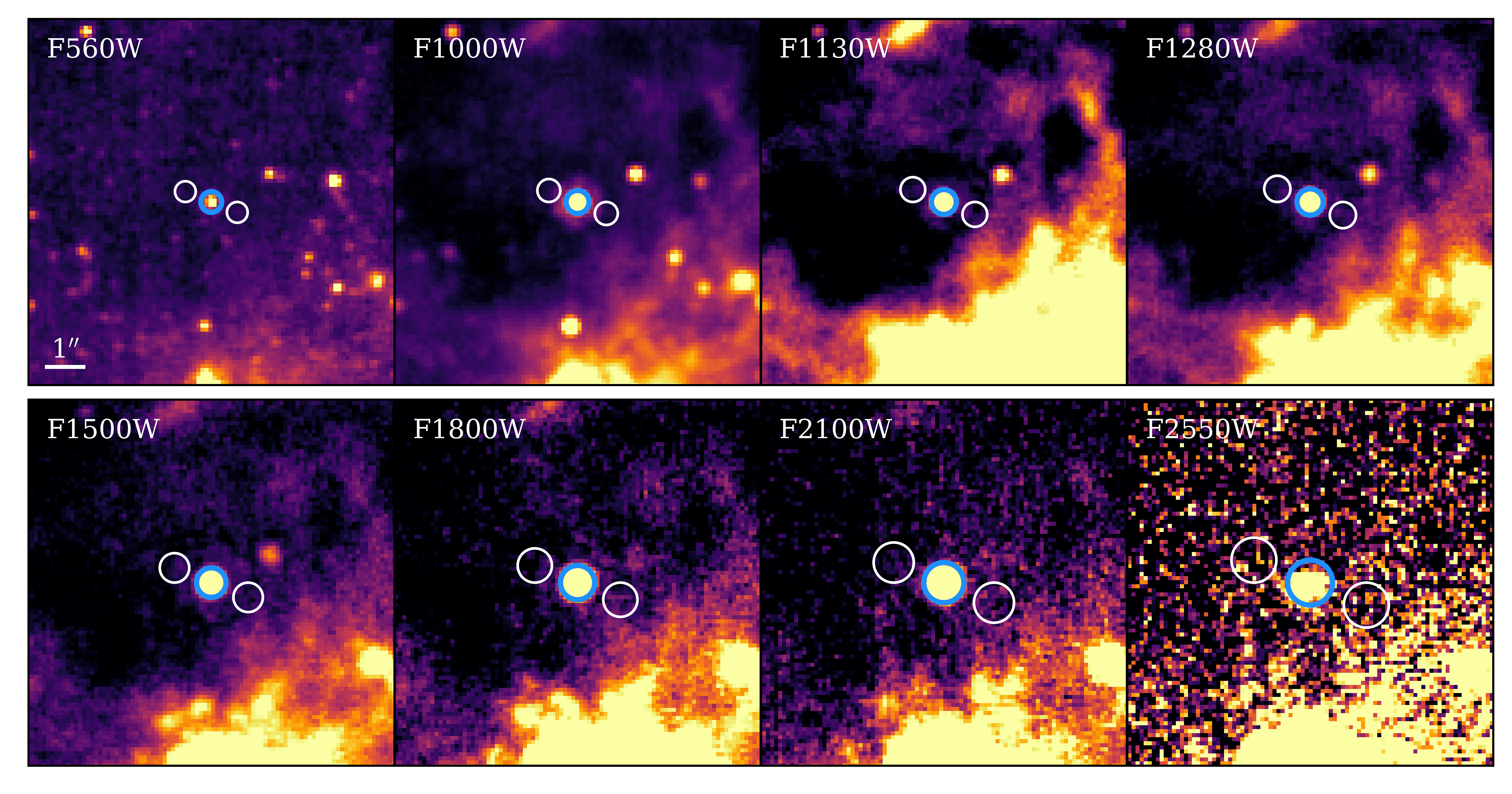}
    \caption{Aperture placement on Cycle 1 JWST/MIRI observations for the two aperture background photometry method. The central blue circle is the science aperture. The two white circles on either side of the target are the apertures used for the background subtraction. We utilized the same aperture locations on the sky for the Cycle 2 observations. Background aperture locations were chosen to avoid diffraction spikes in both the Cycle 1 and Cycle 2 observations.}
    \label{fig:apphot}
\end{figure}

Aperture photometry methods were unable to reproduce the flux values measured by PSF photometry methods. 
Our initial photometry of SN~2017eaw was done using an aperture photometry method similar to that used in \citet{Hosseinzadeh23}, on the Cycle 1 and 2 Level 3 I2D images of SN~2017eaw. The science and background apertures for a selection of the images for SN~2017eaw is shown in Figure \ref{fig:apphot}. 
We choose the science aperture to enclose 60\% of the light from the source based on the JWST/MIRI aperture correction (version 0014, in flight pedigree of 2022-05-25 to 2024-06-02). 
Background subtraction is done using the average of two circular regions on either side of the aperture. We find that in the case of SN~2017eaw the diffraction spikes are minimal enough that using an annulus for background subtraction produces photometry which is consistent with that from the two circular aperture method. However, given the diffraction spikes and surroundings of the comparison stars in the field, we use two circular apertures for consistency. The location of the two background regions were chosen to avoid diffraction spikes while remaining close to the science aperture. The background apertures are chosen to be on the same region of the sky for all exposures of the target. However, as shown in Figure \ref{fig:ap_vs_psf} (green circles), this methodology results in fluxes that are systematically higher than those published in \citet{Shahbandeh23}.

Given the discrepancy between the aperture photometry and the \citet{Shahbandeh23} values, in order to cross check our methodology we utilize a separate aperture photometry code originally designed to do photometry on NIRCAM high-redshift galaxies described in \citet{Endsley23}, which was adapted to allow for aperture photometry on MIRI images. First, SEP \citep[the python library for Source Extraction and Photometry;][]{Sextractor, SEP} is run on the F2100W image to choose an elliptical aperture which encloses $>90$\% of the flux and use this aperture size for all filters. We choose F2100W since it is the reddest filter with a high signal-to-noise point source at the location of SN~2017eaw, and therefore has the largest PSF. Second, SEP is run on all filters to mask out all objects in the field. This requires a background subtracted image. Given the complex background of the image, we opt to create 10"x10" stamps centered around the SN position and measure the spatially varying background using SEP. This background is then subtracted from the image stamp. Third, we randomly place 50 apertures on the background subtracted image and measured their fluxes. The standard deviation in these measurements is the error in our photometry and the median value is subtracted off the final photometry to account for higher order background fluctuations. Finally, we apply an aperture correction determined by dropping the elliptical aperture used on the WebbPSF models \citep{WebbPSF1, WebbPSF2} for each filter and calculating the amount of total flux enclosed within the aperture. 

As shown in Figure \ref{fig:ap_vs_psf} (blue stars), the random aperture background method also yields flux values higher than those reported in \citet{Shahbandeh23}. However, because of how photometric errors are accounted for in the random aperture background method, the error on these measurements are large. Therefore the photometry from this method is roughly consistent with both the previously discussed aperture photometry method and the \citet{Shahbandeh23} PSF photometry.  

Due to the discrepancy between the aperture and PSF photometry regardless of methodology, and the existence of published JWST/MIRI PSF photometry of SN~2017eaw, we opt to report only the PSF photometry discussed in Section \ref{sec:jwstphot} in this work. {We note that \citet{Shahbandeh23} subtract a background image, made by taking the sigma-clipped average of the individual dithers, from the level 2 images before doing PSF photometry. We opt to not do this since \texttt{space\textunderscore phot} fits the background alongside the PSF model and generally our photometry is consistent with \citet{Shahbandeh23}.}
Importantly, we do find that, regardless of the photometric method, the total mid-infrared flux of SN~2017eaw has decreased from Cycle 1 to Cycle 2. We caution that the aperture and PSF photometry of JWST/MIRI data may not be consistent with each other and recommend using similar methods as reported in previous publications. This discrepancy may decrease over the course of the JWST mission as MIRI aperture corrections and PSF models continue to be updated.

\section{Dust Modeling Equations}\label{sec:dusteqs}

In Section \ref{sec:dustmod}, we discuss the fitting methods for the optically thin, dusty sphere, and dusty shell models. Here we present the luminosity equations used for all three of these dust models.

\subsection{Optically Thin Model}
For the optically thin case, we model the input luminosity as two components of dust, with temperatures $T_\mathrm{hot}$ and $T_\mathrm{cool}$ and masses $M_\mathrm{hot}$ and $M_\mathrm{cool}$. We note that in the optically thin case, R$_\mathrm{dust}$ is not well constrained. For each component, the input luminosity is set by:
\begin{equation}\label{eq:L_single}
    L_{\nu,\mathrm{dust}} = 4\pi \kappa_{\nu} M_\mathrm{dust} B_\nu(T_\mathrm{dust}),
\end{equation}
where B$_\nu$(T) is the Planck function and $\kappa_{\nu}$ is the frequency-dependent opacity of the dust component. We calculate $\kappa_{\nu}$ from the absorption efficiency $Q_{\nu}$, particle density $\rho_\mathrm{part}$, and particle size $a$:
\begin{equation}
    \kappa_{\nu} = \frac{3 Q_{\nu}}{4 a \rho_\mathrm{part}}.
\end{equation}
In the optically thin case, the dust will not self attenuate so the total luminosity is just:
\begin{equation}\label{eq:L_thin}
    L_{\nu,\mathrm{thin}} = 4\pi \kappa_{\nu} [M_\mathrm{hot} B_\nu(T_\mathrm{hot}) + M_\mathrm{cool} B_\nu(T_\mathrm{cool})],
\end{equation}

\subsection{Dusty Sphere Model}
In the dusty sphere case, we assume a sphere of dust with total mass $M_\mathrm{hot} + M_\mathrm{cool}$ and two temperature components ($T_\mathrm{hot}$ and $T_\mathrm{cool}$) inside a radius $R_\mathrm{outer}$ with an optical depth $\tau_\nu>0$. This model is geometrically similar to the dusty sphere model used in \citet{Shahbandeh23}. The luminosity of the dusty sphere is extinguished according to the escape probability from \citet[][Appendix 2]{Osterbrock89}: 
\begin{equation}\label{eq:escprob}
    P_\mathrm{esc} = \frac{3}{4\tau_\nu} \left[ 1 - \frac{1}{2\tau_\nu^2} + \left( \frac{1}{\tau_\nu} + \frac{1}{2\tau_\nu^2} \right) e^{-2\tau_\nu} \right],
\end{equation}
here the frequency dependent optical depth ($\tau_\nu$) to the center of a dust shell with bulk density $\rho_\mathrm{bulk}$ is: 
\begin{equation}\label{eq:tau_base}
\tau_\nu = \kappa_\nu \rho_\mathrm{bulk} R_\mathrm{dust} = \frac{3 \kappa_\nu M_\mathrm{dust}}{4 \pi R_\mathrm{dust}^2}.
\end{equation}
For the dusty sphere model $\tau_\nu$ is:
\begin{equation}
\tau_\nu = \frac{3 \kappa_\nu (M_\mathrm{hot}+M_\mathrm{cool})}{4 \pi R_\mathrm{outer}^2}.
\end{equation}

Thus the full SED for the dusty sphere is modeled by:
\begin{eqnarray}\label{eq:L_sphere}
    L_{\nu, \mathrm{sphere}} = \frac{4 \pi^{2} R_\mathrm{outer}^2}{M_\mathrm{hot}+M_\mathrm{cool}} \left[ M_\mathrm{hot} B_\nu(T_{hot}) + M_\mathrm{cool} B_\nu(T_{cool})\right] \nonumber \\ \times \left[ 1 - \frac{1}{2\tau_{\nu, \mathrm{w}}^2} + \left( \frac{1}{\tau_{\nu, \mathrm{w}}} + \frac{1}{2\tau_{\nu, \mathrm{w}}^2} \right) e^{-2\tau_{\nu, \mathrm{w}}} \right].
\end{eqnarray}

\subsection{Dusty Shell Model}
We also fit a dusty shell model, with total dust mass $M_\mathrm{hot} + M_\mathrm{cool}$, inner radius $R_\mathrm{inner}$ and outer radius $R_\mathrm{outer}$, to the SED. For the dusty shell, the frequency dependent optical depth (analogous to Equation \ref{eq:tau_base}) is:
\begin{equation}
\tau_\nu = \frac{3 \kappa_\nu R_\mathrm{outer}}{4 \pi (R_\mathrm{outer}^3-R_\mathrm{inner}^3)}(M_\mathrm{hot}+M_\mathrm{cool}).
\end{equation}
The escape probability is similarly more complex, as it must take the inner cavity, where $\tau_\nu=0$, into account. We use the escape probability expression worked out in \citet{Dwek2024}:
\begin{equation}\label{eq:escprob_sphere}
    P_\mathrm{esc} = \frac{1}{2\tau_\nu} \left[\frac{f(u, \tau_\nu)}{f_0(u)}\right],
\end{equation}
where
\begin{equation}
    \begin{split}
    f(u, \tau_\nu) = & \int_0^{x_c} \left[ 1 - e^{-2\tau_\nu x} \right] x \, dx \\
    &+ \int_{x_c}^1 \left[ 1 - e^{-2\tau_\nu x \left( 1 - \sqrt{1 - \frac{1 - u^2}{x^2}} \right)} \right] x \, dx,
    \end{split}
\end{equation}
and
\begin{equation}
    \begin{split}
    f_0(u) = & \int_0^{x_c} x^2 \, dx \\
    &+ \int_{x_c}^1 \left[ 1 - \sqrt{1 - \frac{1 - u^2}{x^2}} \right] x^2 \, dx,
    \end{split}
\end{equation}
where $x_c = \sqrt{1-u^2}$, $u = \frac{R_\mathrm{inner}}{R_\mathrm{outer}}$, and $R_\mathrm{inner}$ and $R_\mathrm{outer}$ are the inner and outer radii of the shell, respectively.

Which yields the total dusty shell luminosity:
\begin{eqnarray}\label{eq:L_shell}
    L_{\nu, \mathrm{shell}} =  \frac{2 \pi \kappa_\nu f(u, \tau_\nu)}{\tau_\nu f_0(u)} \left[ M_\mathrm{hot} B_\nu(T_\mathrm{hot}) + M_\mathrm{cool} B_\nu(T_\mathrm{cool})\right].
\end{eqnarray} 

\begin{figure}[b!]
    \centering
    \includegraphics[width=\hsize]{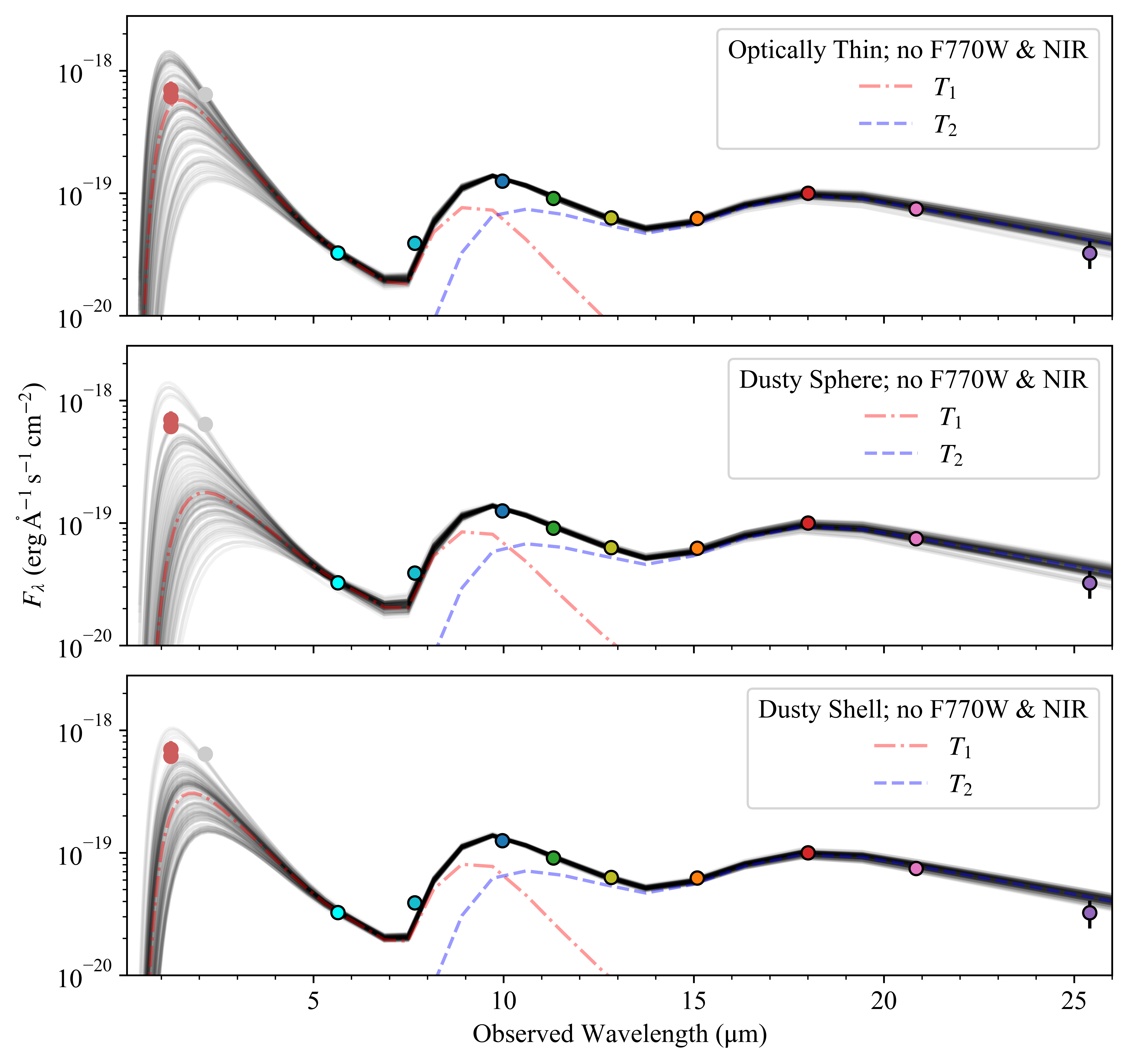}
    \caption{The best-fitting double silicate dust models for the Cycle 2 SEDs when F770W and the NIR bands are excluded. The NIR and F770W fluxes are plotted for comparison. The NIR data in particular is vital to constraining the hot dust component. The observed NIR fluxes are at the high end of the range of fits for the dusty sphere and shell models which assume $\tau_\nu > 0$. }
    \label{fig:No770NIRdustmodels}
\end{figure}

\section{Cycle 2 Dust models excluding F770W and NIR Observations}\label{sec:no770NIR}

There was no NIR or F770W data taken coincident with the Cycle 1 observations of SN~2017eaw. The dust modeling presented here highlights the need for additional constraints of the hot component of the dust SED, especially for younger SNe like SN~2017eaw. Given the lack of constraints on the hot component at 1960 days, we also fit the 2330 day SED with the NIR and F770W photometry excluded so that the SEDs are directly comparable. These results are presented in the furthest right column of Table \ref{tab:dustparameters} and fits are displayed in Figure \ref{fig:No770NIRdustmodels}.

We find that the dust models without the NIR and F770W observations are consistent with the values determined for Cycle 1. The  Cycle 2 models that exclude NIR and F770W photometry tend to favor lower temperatures for the hot components than those found for Cycle 1. Given that the hotter dust component is set by only F560W in these fits, we attribute the decrease in temperature to the decrease in F560W flux. Without NIR observations, it is impossible to track the temperature evolution of the hot dust component.
This uncertainty highlights the need for NIR observations to complement further SNe dust studies. In the case of SN~2017eaw, the majority of the dust mass is in the cool component and the MIRI observations alone are sufficient to suggest that the dust mass of SN~2017eaw did not markedly increase over the course of a year. However, we caution that this may not be the situation for every supernovae, therefore NIR observations are crucial to understanding the dust budget of core-collapse supernovae.

\end{document}